\documentclass[a4paper,12pt]{article}
\usepackage{amsfonts,amsthm,amsmath,amssymb}

\newcommand{\tr}{\operatorname{tr}}

\def\ra{\rangle}
\def\la{\langle}
\def\a{\alpha}
\def\b{\beta}

\def\g{\gamma}

\def\s{\sigma}

\def\d{\partial}

\def\C{\mathbb{C}}

\def\bX{ { \bf{X }}  }

\newcommand{\cL}{\mathcal L}

\newcommand{\cN}{\mathcal N}
\newcommand{\cO}{\mathcal O}

\newcommand{\cR}{\mathcal R}

\newcommand{\cW}{\mathcal W}

\newcommand{\be}{\begin{equation}}
\newcommand{\bea}{\begin{eqnarray}}

\newcommand{\ee}{\end{equation}}
\newcommand{\eea}{\end{eqnarray}}

\newcommand{\nn}{\nonumber}

\def\bu{ {\bf{u}} }  
\def\bl{ {\bf{l}} }

\begin{document}

{}~
{}~
\hbox{QMUL-PH-08-13}
\break

\vskip .6cm

\centerline{{\LARGE \bf  Enhanced symmetries of gauge theory 
 }} 
\centerline{{\LARGE \bf  and resolving the spectrum of  local operators }}

\medskip

\vspace*{4.0ex}

\centerline{ {\large \bf Yusuke Kimura}\footnote{y.kimura@qmul.ac.uk}
{ \bf  and }{\large \bf Sanjaye Ramgoolam}\footnote{s.ramgoolam@qmul.ac.uk}  } 
\vspace*{4.0ex}
\begin{center}
{\large Centre for Research in String Theory, \\
 Department of Physics, \\
Queen Mary, University of London\\
Mile End Road\\
London E1 4NS UK\\
}
\end{center}

\vspace*{5.0ex}

\centerline{\bf Abstract} \bigskip

Enhanced global non-abelian symmetries at zero coupling in Yang Mills theory 
play an important role in diagonalising the two-point functions of 
multi-matrix operators. 
Generalised Casimirs constructed 
from the iterated commutator action of these 
 enhanced symmetries resolve all  the multiplicity labels of the 
bases of matrix operators which diagonalise the two-point function. 
For the case of $U ( N )$ gauge theory with  a single complex  matrix
in the adjoint of the gauge group   we have a $ U(N)^{\times 4 } $ 
global symmetry of the scaling operator at zero coupling. 
Different choices of commuting sets of Casimirs, for the case of 
a complex matrix, lead to the restricted Schur basis 
 previously studied in connection with string excitations of
 giant gravitons and the Brauer basis studied in connection with 
brane-anti-brane systems.  
More generally these remarks can be extended to the
diagonalisation for any global symmetry group $ G$. Schur-Weyl duality 
plays a central role in connecting the enhanced symmetries and the 
diagonal bases.

\newpage
\tableofcontents

\setcounter{footnote}{0} 

%------------------------

\section{Introduction }

Local gauge invariant operators in a conformal field theory
correspond to physical states. Understanding the spectrum of 
states from the point of view of the CFT yields information 
about spacetime physics via the AdS/CFT duality \cite{malda,gkp,wit}. 
The two-point functions of the CFT give an inner product on 
the states. The diagonalisation of this inner product
is a valuable tool in the detailed understanding of spacetime 
physics. States created by holomorphic  multi-traces of  
one complex matrix are half-BPS. The diagonalisation 
in this sector  has interesting connections 
with free fermions \cite{cjr,ber} , giant gravitons \cite{mst}, LLM geometries
\cite{llm} and black hole physics \cite{bbjs} in the context 
of  $U(N)$ gauge theory with  $ \cN =4 $ supersymmetry 
and its string dual in $AdS_5 \times S^5 $. 

Recent progress on the diagonalisation of the 
two-point functions  of gauge invariant multi-matrix 
operators  \cite{kr,bhr,bcd,bhr2}
has generalised  earlier work on the holomorphic sector 
of a single complex matrix \cite{cjr,cr}.
In this paper, we will explain how the labels appearing in the 
diagonal bases are related to Casimirs constructed 
from Noetherian symmetries in the zero coupling limit. 
The construction of these diagonal bases heavily 
uses symmetric group data, or related finite algebras such as  
Brauer algebras.  The Noetherian symmetries are 
unitary Lie group  symmetries, e.g products of $U(N)$  or 
$U(N^2)$. Schur-Weyl duality explains the relation 
between these unitary symmetries and the symmetric group 
construction of the diagonal bases.

The first case we consider is the sector of 
one complex matrix. In this case we can use two 
diagonalisation methods :  the Brauer Basis
 \cite{kr} and the restricted Schur basis \cite{bcd}.
 The second case we will consider is the 
case of the holomorphic sector of 
$M$  complex matrices.  We also sketch the application 
of enhanced symmetries in the case of the diagonalisation 
with general global symmetry $G$ given in \cite{bhr2}. 

\subsection{Schur-Weyl duality and enhanced symmetries } 
 
The diagonal basis for the holomorphic sector of 
one complex matrix $X$, which is relevant for the half-BPS 
representations of $ \cN =4 $ SYM, is labelled by 
Young diagrams $R$ of $U(N)$, i.e those with first  column 
no longer than $N$.  We have $ \cO_R = \chi_R ( X ) $. 
For operators made from 
$n$ fields $X$, the Young diagrams have $n$ boxes. 
This means that $R$ is a partition of $n$ which we write 
as $ R \vdash n  $ and the constraint on  the columns is expressed 
as $ c_1(R ) \le N$. 
These Young diagrams are also associated with representations 
of the symmetric group $S_n$ of all permutations of $n$ 
objects. To understand the role of $U(N) $ and $S_n$ 
it is useful to view the $ X  $ as an operator acting on 
an $N$-dimensional vector space $V$. It can be extended to 
${ \bf X } = X \otimes X \cdots \otimes X $ acting on $V^{\otimes n } $. 
The basic reason for the appearance of $ S_n$ is that 
elements $ \sigma $ of the symmetric group organise the 
space of multi-traces. The multi-trace operators can be written as 
$ tr_n ( \sigma \bX ) $ for $ \sigma \in S_n $ with $tr_n$ 
denoting a trace in $V^{\otimes n } $. The role of $R \vdash n $ with 
$c_1( R ) \le N $ can be understood by thinking about the 
decomposition of $ V^{ \otimes n } $ in terms of $ U(N)\times S_n$
\bea\label{basicsw}  
 V^{ \otimes  n } = \bigoplus_{ R } V_{ R }^{ U(N) }  \otimes V_R^{ S_n }   
\eea   
This equation is called Schur-Weyl duality and follows from the 
fact that the algebra of operators commuting with $ U(N)$ in
$V^{ \otimes  n }$ is $ \C ( S_n) $, the group algebra of 
$S_n$. Conversely the algebra commuting with $S_n$ 
is the universal enveloping algebra of $ U(N)$. For a review 
of results in two and four dimensional Yang Mills which rely 
on (\ref{basicsw}) and its generalisations, see \cite{swreview}.  

A critical physicist might argue that the above account 
raises a small mystery. The $U(N) $ gauge symmetry 
of  Yang Mills is not a dynamical symmetry acting on 
the physical spectrum. 
On the other hand a symmetry that organises operators 
is a dynamical symmetry. Indeed operators in conformal 
field theory correspond to states and an algebra that organises 
the operators organises the states. The resolution 
is that, at zero coupling the global part of the $U(N)$ gauge symmetry is 
part of a bigger symmetry which includes $U(N) \times U(N)$ 
acting separately on the lower and upper indices of $ X^i_j$. 
This is explained in more detail in section \ref{sec:onecomplex}. 
This enhanced symmetry commutes with the classical dilatation 
operator and hence leaves unchanged the space of operators 
made from a fixed number of $X$. It turns out that the enhanced 
symmetries can be used to construct Casimirs which act on 
$ \cO_R $ to give eigenvalues $ C_2( R ) , C_3(R) \cdots $.
 This remark was essentially 
already contained in  \cite{cjr} where it was discussed in the 
context of a reduced matrix model on $S^3 \times R $ where the 
complex field has a mass term. Here we will not  do the
dimensional reduction and   will discuss global symmetries of 
the four dimensional theory.  The Casimirs constructed in 
the context of the holomorphic $X$ sector  have  been viewed 
as charges in spacetime \cite{integinfo}. Viewing these charges as 
physical observables in the dual spacetime 
can be used to justify interest in  the basis $ \cO_R $.
From a technical point of view, it is worth observing that 
states which have distinct eigenvalues of a complete set of 
Casimirs form an orthogonal basis by a  standard quantum mechanics 
argument. 
We may view the fact these diagonal bases form eigenstates 
of the Casimirs as evidence that they are useful for physics. 
In the  case of the  more general diagonalisations
involving multiple matrices and global symmetries,  
 the diagonal bases found recently use, along with 
Young diagram labels, some more subtle group theoretic 
labels from the world of symmetric groups and Brauer algebras. 
We find that Schur Weyl duality is a valuable guide 
which helps anticipate the type of Casimirs, 
to be constructed from enhanced symmetries of the Yang Mills 
Lagrangian at $ g_{YM}^2 = 0 $, relevant in these more general cases.

One immediate  consequence of Schur-Weyl (SW)  duality (\ref{basicsw})
is that Casimirs constructed from $ U(N) $ generators 
can be expressed in terms of elements of $ \C ( S_n ) $
(see equation (\ref{casUS})). This intimate connection 
between Casimirs and SW duality can be generalised in several directions.  
 The classical Schur-Weyl duality above 
is a special case of a more general theorem 
called the double commutant theorem, which gives a similar decomposition 
for any space $\cW$ under the action of an algebra $A$
\bea 
 \cW = \bigoplus_{ \Lambda  }  V_{ \Lambda }^{ A } \otimes V_{ \Lambda }^{ Com ( A ) } 
\eea 
In the above $ Com ( A ) $ is the commutant of $A$,  
$V_{ \Lambda}^{ A }$ is an irreducible representation (irrep) 
of $A$, $ V_{ \Lambda}^{ Com ( A ) }$ is an irrep of the $Com(A) $. 
This implies that the multiplicity of each irrep of $A$ is 
given by the dimension of a corresponding irrep of $Com ( A ) $. 
For details on the double commutant theorem (also called the 
double centraliser theorem),  see section 1
of \cite{halverson} or textbooks such as \cite{goodwall}. 
The following fact from double commutant theory will be useful. 
Suppose $A$ acts on $ \cW $ and has commutant $ Com( A )$. 
Suppose further that $A$ has a subalgebra $B$. Clearly 
$ Com ( A ) $ is a subalgebra of $Com ( B)$. In this case 
suppose $ V_{ \mu }^{ B } $ appears in the sub-algebra decomposition 
 of $V_{ \lambda }^{ A }  $ with multiplicity $g_{ \lambda \mu } $. 
And further suppose that the decomposition of  $ V_{ \mu }^{ Com ( B ) } $
in terms of the sub-algebra $Com ( A ) $ contains $ V_{ \lambda }^{Com(A)}  $ 
with multiplicity $ g_{ \mu \lambda }^{\prime}  $. Then 
the useful result is $ g_{ \lambda \mu } =   g_{ \mu \lambda }^{\prime}$. 
This implies that various group theoretic  multiplicities 
defined in the world of symmetric groups have a dual meaning 
in the world of unitary groups and vice versa.

We will make extensive use of the above concept. 
In sections \ref{sec:ehsBrauer} and  \ref{sec:PfsBrauerXX*} 
we will be considering a diagonal basis of operators \cite{kr}  
in the sector of $ X , X^{*} $ given in terms of 
the Brauer algebras $B_N ( m, n) $ and its reduction 
to the subalgebra $ \C ( S_m ) \times \C ( S_n ) $. 
The SW dual of $ B_N ( m, n) $ is  $U(N)$ acting on 
$V^{\otimes m }  \otimes \bar{V}^{\otimes n  } $. So we expect the 
Casimirs to involve multiple copies of $U(N)$. All the relevant 
$U(N)$ symmetries of the classical scaling operator 
 are found in  section \ref{sec:onecomplex}. A different combination 
of the same $ U(N)$ symmetries is used to construct Casimirs 
which resolve the labels on the restricted Schur basis of 
operators given in \cite{bcd} which involve the reduction 
of $ \C ( S_{m+n } )  $ to $ \C ( S_m ) \times \C ( S_n ) $.
 This is done in sections 
\ref{ehsrestschu} and \ref{PfssymmXX*} . 
Another case considered is the sector of $M$ complex matrices 
where a diagonal basis of holomorphic operators was given 
in \cite{bhr} which is $U(M)$ covariant. The labels on the 
basis include a Clebsch multiplicity   $ \tau $ which runs 
over the number of times the irrep $\Lambda$ of $S_n $ appears    
in the tensor product (sometimes called inner tensor product )
  $ R \otimes R $ of the irrep $ R $ of 
$ S_n   $. Now the tensor product  $ R \otimes R $
is a representation of $ S_n \times S_n $ and the Clebsch 
reduction problem is equivalently 
the problem of decomposing the  representations 
of the product group in terms of the diagonal $S_n$ subgroup. 
This inner tensor product problem of symmetric groups is 
known ( see e.g \cite{hassbut}) 
 to be related by Schur Weyl duality to the embedding of 
$U(N) \times U(N)$ inside $U(N^2)$ (in fact more generally
to the embedding of $ U(M) \times U(N)$ to $U(MN)$).  
 We will describe the 
relevant $U(N^2)$ as symmetries constructed from  the free field 
lagrangian in section \ref{sec:enhanced}. 
In  sections \ref{sec:umsnI} and \ref{sec:umsnII} 
we will show how the generators of $U(N^2)$ along with $U(N)$ can be 
used to construct the invariant generalised Casimirs which 
distinguish the operators with different values of $ \tau $. 

In section \ref{sec:examples} we give computations 
of some of  the eigenvalues 
of these generalised Casimirs.

\section{Enhanced Symmetries in  zero coupling gauge theories  } 
\label{sec:enhanced}

Zero coupling  $U(N)$ gauge theories have a large global symmetry group. 
The key features are best described in the simple example of 
a complex scalar transforming in the adjoint of the gauge group. 
We first describe a $U(N)^{ \times 4 } $ symmetry whose 
diagonal subgroup  acts  the gauge transformation. The diagonal 
of course leaves all the gauge invariant operators unchanged. 
But the full group acts non-trivially on the spectrum of gauge 
invariant states. Then we describe a 
$U(N^2)\times U(N^2) $ symmetry which contains the $U(N)^{\times 4 } $. 
This symmetry essentially follows from the fact that we have 
$N^2$ free fields.

\subsection{One complex Matrix : $U(N)^{ \times 4 } $ } 
\label{sec:onecomplex}

In $ \cN =4 $ SYM, the highest weight states are in 
1-1 correspondence with local operators constructed from 
 holomorphic multi-trace combinations of a single complex matrix
$ X = \phi_1 + i \phi_2 $, where $\phi_1 , \phi_2 $ are two of the 
six hermitian matrices in $\cN =4 $ SYM.  

The action for the complex scalar coupled to Yang Mills is 
\bea\label{act} 
&&   { 1 \over 2 }  \int  tr ~ F_{\mu \nu } F^{\mu \nu } +
 tr ~ D_{\mu } X D^{\mu } 
X^{\dagger} 
\eea
We are interested in gauge invariant operators constructed 
from the scalar and their correlators at zero coupling.
For these computations of correlators the relevant part of the action is 
\bea 
\int tr ~ D_{\mu } X D^{\mu } X^{\dagger}
\eea 
Working in the $A_0 =0 $ gauge we can separate out the 
part of the action containing $A_0$, which acts 
as a Lagrange multiplier for the Gauss Law Constraint \cite{itzub}. 
The matter coupling of $A_0 $ coming from $  D_{0 } X D_{0} X^{\dagger}$ is 
\bea 
\int tr ( A_0 X^{\dagger}  \d_0 X + A_0 X \d_0 X^{\dagger} - A_0\d_0 
X X^{\dagger}  - A_0 \partial_0 X^{\dagger} X  )      
\eea  
In canonical quantisation the Gauss Law matrix operator 
\bea 
{\cal G} =   X^{\dagger}  \d_0 X +  X \d_0 X^{\dagger} - \d_0 
X X^{\dagger}  -  \partial_0 X^{\dagger} X
\label{gausslaw}
\eea 
 generates time independent gauge transformations. We will in fact 
be interested in the global transformations which are also 
independent of time, and are generated by 
\bea 
G^i_j  =-i \int d^3 x ~~ { {\cal G}^i_j   } 
\eea

The momenta conjugate to $X^i_j $ and $ X^{\dagger i }_j $  
are  
\bea 
&& { \partial L \over \partial  \partial_0 X ^i_j }=
   \partial_0 {X^{\dagger} }^j_i =  \Pi_{ X^i_j}   \cr 
&&  { \partial L \over \partial    \partial_0 X^{\dagger i}_j }   =
   \partial_0 X^j_i =  \Pi_{ {X^{\dagger}}^i_j }   
\eea 
The canonical commutation relations are 
\bea\label{cancom}  
&&  [ (\Pi_{X})^{j}_i ( x ) , X^p_q ( 0)  ] = i ~ \delta_i^p ~  \delta^j_q 
 ~ \delta^3 ( x )   \cr 
&&  [  ( \Pi_{X^{\dagger} })^{j}_i ( x )  ,  X^{\dagger p }_q ( 0)  ] 
      = i  ~ \delta_i^p ~  \delta^j_q  ~  \delta^3 ( x )
\eea

Corresponding to the four terms in (\ref{gausslaw})
we will write the Gauss Law operator as 
\bea 
G = G_1 + G_2 + G_3 + G_4 
\eea 
and they 
 act as follows 
\bea\label{epsact0}  
[   G_{1}{}_{j}^i ,  X^{\dagger}{}^{p}_q  ] = 
 [ -i \int d^{3}x ( X^{\dagger}  \d_0 X )^i_j (x) , 
 X^{\dagger}{}^{p}_q  ] &=&     
 \delta_j^p  X^{\dagger}{}^{i}_q  \cr 
 [   G_{2}{}_{j}^i ,  X^p_q  ] =  
 [   -i \int d^{3}x ( X \d_0 X^{\dagger} )^i_j(x) , X^{ p}_q ]  &=&
      \delta^p_j X^i_q \cr  
 [    G_{3}{}_{j}^i ,  X^{\dagger p}_q  ] = 
 [ i     \int d^{3}x ( \d_0 X X^{\dagger} )^i_j(x), X^{\dagger}{}^{ p}_q ] & =&
    - \delta^i_q
X^{\dagger}{}^{p}_j \cr 
  [    G_{4}{}_{j}^i ,  X^p_q  ]  =  
 [ i\ \int d^{3}x  ( \partial_0 X^{\dagger} X )^i_j (x), X^{p}_q ] 
    & = &   -   \delta^i_q X^p_j 
\eea 
Equivalently we can express the commutation relations as follows 
\bea\label{epsact}  
&&[   tr(\epsilon_{1}G_{1}) ,  X^{\dagger}{}^ {p}_q  ] = 
     (\epsilon_{1}X^{\dagger})^{ p}_q  \cr 
&&[   tr(\epsilon_{2}G_{2}) ,  X^{ p}_q  ] = 
     (\epsilon_{2}X)^{p}_q  \cr 
&&[   tr(\epsilon_{3}G_{3}) ,  X^{\dagger}{}^{p}_q  ] = 
     - (X^{\dagger}\epsilon_{3})^{p}_q  \cr 
&&[   tr(\epsilon_{4}G_{4}) ,  X^{ p}_q  ] = 
    - (X\epsilon_{4})^{p}_q  
\eea 
The commutation relations of the $ G_{a} $ ( $ a = 1 \cdots 4 $ )
are those of the generators of $U(N)^{\times 4 } $ 
\bea\label{UNcomrels}  
[  ( G_{a} ) ^{ i}_{j} , (G_{b})^{ k}_{l} ] =
 \delta_{ab}  ( (G_{a})^i_l \delta_j^k
  - (G_{a})^{k}_j \delta^i_l   )   
\eea 
Exponentiating these actions we get the action of four 
copies of $ U(N)$, respectively  
\bea\label{unact}  
&& X^{\dagger} \rightarrow U_1 X^{\dagger} \cr 
&& X \rightarrow U_2 X \cr 
&& X^{\dagger}  \rightarrow X^{\dagger}  U_3^{\dagger} \cr 
&& X \rightarrow X U_4^{\dagger}  
\eea 
Given these actions it is natural to rename 
\bea 
&& G_1 = G(\cL,X^{\dagger})  \cr 
&& G_2 = G ( \cL , X) \cr  
&& G_3 = G ( \cR , X^{\dagger} )  \cr 
&& G_4 = G ( \cR , X )  
\eea 
where $\cL$ denotes left action and $\cR$ denotes right action. 
When we set $ \epsilon_1 = \epsilon_2 = \epsilon_3 = \epsilon_4 $ 
we have the adjoint action on $ X , X^{\dagger} $ which is the 
global gauge symmetry. It constrains local operators to be 
constructed from traces. Note however that setting 
$ \epsilon_2 = \epsilon_3 $ we have $ \epsilon_2  ( G_2 + G_3 )  $ 
which is a symmetry of the Lagrangian. Likewise setting
 $ \epsilon_1 = \epsilon_4 $ 
we have $   \epsilon_1  ( G_1 + G_4 )  $ which is also a symmetry. 
In fact, for the comparison to the global time Hamiltonian 
of AdS, we are interested in the scaling operator
\bea 
L_0 = \int d^{3}x\left(X^i_j ( \Pi_X )^j_i 
+  X^{\dagger i}_j ( \Pi_{X^\dagger } )^j_i \right)
    = \int d^{3}x ~ tr\left( X \Pi_X + X^{\dagger } \Pi_{ X^ { \dagger }  }  
    \right)
\eea 
By using (\ref{cancom}) it is easy to check that 
\bea 
&& \bigl [ G_{a } ,  \int d^{3}x ~ tr  \left( X \Pi_X  \right) \bigr ] = 0  
\quad \hbox{ for } a = 1 \cdots 4 \cr 
&& \bigl[  G_{a} , \int d^{3}x ~ tr  
\left ( X^{\dagger } \Pi_{ X^ { \dagger } }  \right)  \bigr] = 0  \quad \hbox{ for } 
a = 1 \cdots 4
\eea 
 
So we have, in the zero coupling limit, a symmetry 
of $ U ( N ) \times U ( N ) \times U ( N ) \times U (N ) $ 
of the scaling operator. The diagonal of these actions 
is nothing but the gauge group action under which all 
local traced  operators are going to be invariant. However we 
are also interested in the full $ U(N)^{ \times 4 } $ in organising 
the gauge invariant operators into a basis. 
As an aside note that if we take the parameters $ \epsilon^i_j$ 
to be general complex numbers we actually have $GL(N , \C )^4 $, 
whereas to get $ U(N)^{ \times 4 } $ we let $ \epsilon$ be hermitian. 
For our applications  of symmetry groups organising the
diagonal bases for two-point functions, it suffices to use 
unitary groups rather than the complex form, since the 
unitary groups have the same commutants as the general linear groups.

\subsection{$U(N^2) \times U(N^2)$ symmetry }\label{sec:UNsq} 
 
The free  action for a complex matrix in fact has
a $U(N^2)$ symmetry. This is clear by writing 
\bea 
\int d^{4} x ~ tr ( \partial_0 X \partial_0 X^{\dagger} ) 
= \int d^{4}x  ~  \partial_0 X^i_j  \partial_0 X^{ i *  }_j  
= \int d^{4}x   ~ \partial_0 X_A   \partial_0 X^{*}_A  
\eea 
where $ A = ( i , j ) $ is a composite index, taking 
$N^2$ values as $ i , j $ run over $N$ values. 
 In canonical quantisation  we have generators 
\bea 
E^{ jk}_{lm} (  X )  =  -i\int d^3 x    X^k_l \Pi_{ { X^m_j }  }  
\eea 
They  transform the $X $ as 
\bea 
[ E^{ jk}_{lm} ,  X^{p}_{q} ]  = \delta^{p}_{m} \delta^{j}_{q} X^k_l 
\eea 
 Likewise we have a $U(N^2) $ 
symmetry for the $ X^{\dagger} $. 
\bea 
E^{ jk}_{lm} (  X^{\dagger}  )  = -i \int d^3 x    X^{ \dagger k }_l 
\Pi_{ { X^{\dagger m}_j }  }  
\eea 
 One checks that these are generators of a  
$ U(N^2) \times U(N^2)$ algebra which leaves the 
classical scaling operator  $ tr ( X \Pi_{ X} ) $ invariant.  
Each $U(N^2)$ has a $ U(N) \times U(N)$ symmetry, 
so the $ U(N)^{ \times 4 } $ is a subgroup of 
the $ U(N^2)\times U(N^2) $.

For $M$ complex matrices, $ X_1, X_2, \cdots ,X_M$ 
we have $m$ copies $ (  U(N^2)  \times U(N^2)  )^{\times M } $ 
   generated by 
$ ( E^{ jk}_{lm} ( X_1 ) ,E^{ jk}_{lm} ( X_1^{\dagger}  ) ) ,
    \cdots ,  ( E^{ jk}_{lm} ( X_M )  ,E^{ jk}_{lm} ( X_M^{\dagger} )  ) $. 
Defining in this  case 
\bea 
 E^{jk}_{lm}  =   E^{jk}_{lm} ( X_1 ) + \cdots +  E^{ jk}_{lm} ( X_M ) 
\eea 
This will act as 
\bea 
 [ E^{jk}_{lm} ,  (X_{a})^p_q  ]  = \delta^{p}_{m} \delta^{j}_{q} (X_a)^k_l
\eea 
for any $a= 1 \cdots M $.

\subsection{Enhanced symmetries of $\cN=4$ SYM  } 
 The full SYM has a product group symmetry  
$ U(N)^{\times 4 } \times U(N)^{\times 4 } \cdots \times U(N)^{\times 4 }  $ 
consisting of $8$ factors, of which $4$ act on 
bosonic fields and $4$ act on fermionic fields. 
There are $3$ bosonic $U(N)^{\times 4} $ coming from 3 complex scalars. 
There is $1$ bosonic $U(N)^{\times 4}$ for the gauge field, which is 
easiest  to describe in the light-cone gauge where the 
kinetic term takes the form $ \int \tr \partial_+
 A_z \partial_- A_{\bar z } $ and a discussion very similar to the 
above treatment of a complex scalar field can be done. 
These  four copies of $U(N)^{ \times 4 } $ are subgroups 
of four copies of $ U(N^2) \times U(N^2) $. 
In the fermionic sector, there are four  two-component Weyl fermions 
with kinetic term 
$ i \int tr \partial_m \bar \Psi_a \bar  \sigma^m \Psi_a  $ with 
the flavour index $a$ taking values from $1$ to $4$.  
For each fixed flavour index $a$, we have 
charges
\bea 
   ( G_{ \alpha } ( \cL , \Psi )  )^i_k & = &
  (  \Psi_{\alpha})^i_j \Pi_{ ( \Psi_{\alpha} )^k_j  }  \cr 
  ( G_{ \alpha } ( \cR  , \Psi )  )^i_k & = &
\Pi_{ ( \Psi_{\alpha} )^j_i  } (  \Psi_{\alpha})^j_k
\eea  
As $ \alpha $ takes two values, there are four  commuting copies of 
$U(N)$ for each of the four flavours.
This gives  four copies of  $U(N)^{\times 4} $ 
in the fermionic sector, which are subgroups of $(U(N^2))^{ \times 2 } $

\section{Casimirs of Enhanced symmetries and Diagonal bases 
 } 

\subsection{Enhanced symmetry and Casimirs for Holomorphic sector of $ X $ } 
The operators $ \cO_R = \chi_R ( \bX  ) $ are eigenstates 
of the Casimirs constructed $ tr  ( ( Ad_{ G_{ \cL }} )^n )  $ 
where $ G_{\cL}  = G ( \cL , X ) $.  This remark was contained 
in \cite{cjr} in the slightly different context of 
the   reduction to quantum mechanics. 

Recall some standard group theory of the generators $E^i_j$ of $U(N)$
\bea  
[ E^i_j , E^k_l ] = \delta^k_j E^{i}_l - \delta^i_l E^k_j
\eea
Consider the vectors in the fundamental representation $ V$
\bea  
 ( E )^i_j v^p  = \delta^p_j v^i  
\eea 
On the tensor space $V^{ \otimes n } $ we have the action  
\bea 
&&   ( E )^i_j   (  v^{p_1} \otimes v^{p_2} \cdots \otimes v^{p_n} )   \cr 
&& =  E^i_j  ( v^{p_1} )  \otimes v^{p_2} \otimes \cdots \otimes v^{p_n}  +
  v^{p_1}  \otimes  E^i_j  ( v^{p_2} )  \otimes \cdots \otimes  v^{p_n}  + 
 \cdots +  v^{p_1}  \otimes v^{p_2} \otimes \cdots \otimes E^i_j (  v^{p_n} )
\nn\\  
\eea 
We know that the Casimir will be constant on 
the irreducible subspaces of $ V^{\otimes n } $. 
The decomposition of  $ V^{\otimes n } $ in terms of $U(N) \times S_n$ 
is given by Schur-Weyl duality 
\bea\label{SWduality}  
 V^{\otimes n } = \bigoplus_R V_R^{ U(N) }  \otimes V_R^{S_n}  
\eea 
The projector for $R$ in terms of $S_n$ group theory 
can be given as 
\bea 
p_R =  { d_R \over n ! }  \sum_{\sigma} \chi_R ( \sigma ) \sigma 
\eea 
Hence it follows that 
\bea\label{cas2act} 
 \hat{C}_2  ~  p_R  (  v^{p_1} \otimes v^{p_2} \cdots v^{p_n} ) 
= C_2 ( R ) ~ p_R  (  v^{p_1} \otimes v^{p_2} \cdots v^{p_n} ) 
\eea 
This can be seen to follow from (\ref{SWduality}) and the 
fact that $p_R$ projects the direct sum to a fixed $R$, 
so the eigenvalue is the quadratic $U(N)$ Casimir 
$C_2(R) $. Schur-Weyl duality also gives more information on $ C_2(R)$.

It follows from Schur-Weyl duality (\ref{SWduality})  that the action of 
$\hat{C}_{2}=E^{i}_jE^j_i$ on $ V^{ \otimes n } $, which 
commutes with $ U(N)$, must be expressible in terms 
of central elements of $ \C ( S_n )$.
 Since the Casimir commutes with $U(N)$ 
 and the commutant   of $ U(N)$ in $ V^{ \otimes n } $ 
 is $ S_n$, it follows that the Casimir is in  $ \C ( S_n )$. 
 Since it is constructed from generators of $U(N)$ which 
 commute with $S_n$, this means  that it must be a central 
 element, i.e something in $ \C ( S_n ) $ which 
 commutes with  all elements in $ \C ( S_n ) $. Explicitly, 
\bea\label{casUS}  
\hat{C}_2 & =&  N n + \sum_{ r \ne s } (rs ) \cr 
    & \equiv &  Nn + 2 T_2 
\eea 
We review the derivation in Appendix A. 
The consequence for  eigenvalues is 
\bea\label{C2eigval}  
C_2 ( R ) &=& Nn + 2 { \chi_R ( T_2 ) \over d_R  } \cr 
          &=& Nn +  \sum_{ i } r_i ( r_i - 2 i +1  )
\eea 
Here $d_R $ is the dimension of the symmetric group representation $R$, 
and the last line follows using a standard result on 
the characters of $S_n$. The above equation (\ref{C2eigval}) 
plays an important role in the string theory of two dimensional Yang Mills
\cite{GT}. To get the explicit formula 
(\ref{C2eigval}) for eigenvalues  $C_2(R)$ it is useful 
to use (\ref{casUS}).

The Schur Polynomial operators can be written as 
\bea 
\cO_R \equiv \chi_R ( \bX ) &=& { 1 \over n! } 
\sum_{ \sigma } \chi_R  ( \s ) tr_n ( \s \bX )  \cr 
  & =& { 1 \over d_R } tr_n ( p_R \bX ) \cr 
  & =& {  1 \over d_R }  ( p_R )^J_I   ( \bX )^I_J   
\eea 
The last expression will be useful. Here $I,J$ are multi-indices 
e.g $I = (i_1 , i_2 , \cdots , i_n )$.  
The action of $Ad_{ ( G_{\cL })^i_j } $ on the upper indices of $\bX $
is the same as the action of $E^{i}_{j}$ on $V^{\otimes n }$. Hence
the iterated commutator action is 
\bea 
 [  (G_{\cL } )^i_j ,  [ ( G_{\cL })^j_i  , \cO_R  ]] 
 &=&   Ad_{ ( G_{\cL } )^i_j }Ad_{ ( G_{\cL })^j_i } ~ \cO_R \cr 
 & = &  { 1 \over d_R }  ~ ( p_R )^J_I  ( \hat C_2 )^I_K  ~  ( \bX )^K_J \cr 
 & = &   { 1 \over d_R }  ~ ( \hat{C}_2 p_R )^J_K  ~ ( \bX )^K_J \cr 
 & = & { 1 \over d_R }    ~ C_2 ( R ) ( p_R )^J_K  ~  ( \bX )^K_J \cr
 & = & C_2 ( R )  ~ \cO_R 
\eea 
In the third line we used (\ref{cas2act}).

\subsection{Enhanced symmetry and Casimirs : Brauer Basis for $ X $, 
$ X^{ \dagger} $ } 
\label{sec:ehsBrauer}  

Consider $G_B \equiv  G_2 + G_3 $  which generates 
$ X \rightarrow U  X  , X^{\dagger } \rightarrow X^{\dagger } U^{\dagger}$ 
equivalently 
\bea 
&& X \rightarrow U X  \cr 
&& X^* \rightarrow U^* X^*  
\eea 
On gauge invariant operators $ G_B = - G_1 - G_4 $ 
because $\sum_{a=1}^{4}G_{a}$ generates the adjoint gauge 
transformation $X\rightarrow UXU^{\dagger}$, 
$X^{\dagger}\rightarrow UX^{\dagger}U^{\dagger}$. 
Under the above action, the lower indices of $ X , X^* $ are 
inert while the upper indices transform as $ V \otimes \bar{V} $, 
i.e the fundamental and its complex conjugate. Composite 
operators are built by considering $ m $ copies of $ X $ 
and $ n $ copies of $ X^* $. It is useful to view these 
as operators of the form $ X \otimes X \cdots \otimes X  \otimes X^* \otimes \cdots X^* $ acting on $ V^{ \otimes m } \otimes \bar{V}^{\otimes n } $.   
The Schur Weyl dual of the $U(N)$  action on this space 
is the Brauer algebra $  B_N ( m , n ) $

The Brauer algebra can be
 used to construct a diagonal basis \cite{kr} 
 of 
gauge invariant operators $ \cO^{\g}_{ \a , \b ; i ,  j } \equiv 
 tr_{m,n} ( Q^{\gamma}_{\alpha \beta ; i j } (  \bX \otimes \bX^{* } )) $. 
The label $ \g $ denotes an irrep of $B_N(m,n)$. 
The labels $ \alpha , \beta $ denote irreps. of $ S_m , S_n$ 
respectively. Equivalently $ ( \a , \b ) $ is an irrep of the 
product group $ S_m \times S_n $. We will often use the label 
$A\equiv ( \alpha , \beta ) $ as an abbreviation and 
states will be denoted with $m_A \equiv  ( m_{\a} , m_{ \b} )$. 
The indices $i,j$ here each run over the 
 mutiplicity  of $A$ in the restriction 
of the irrep $ \g$ of $B_N (m,n )$ to the irrep $A$ of the subalgebra 
$ \C ( S_{m} \times S_n ) $.  Note we also often use $i,j$ among 
$U(N)$ indices :  viewing the formulae in context should  resolve
 any possible confusion. 
More on this  basis of operators 
in the Appendix \ref{sec:SymmetricBranchingOperator}.

We will show the following iterated commutator actions 
\bea\label{bbascaseqs}  
   tr  ( Ad_{ G_B}^2 )~ \cO^{\g}_{ \a , \b ; i,  j } 
  &=~~~~~~~~~ Ad_{ (G_B)^i_j }  Ad_{ (G_B)^j_i } ~   \cO^{\g}_{ \a , \b ; i,  j } &= 
C_2 ( \gamma ) ~ \cO^{\g}_{ \a , \b ; i,  j } \cr 
 tr ( Ad_{ G_2}^2) ~\cO^{\g}_{ \a , \b ; i,  j } &=~~~~~~~~~~
 Ad_{ (G_2)^i_j} Ad_{ (G_2)^j_i}    ~ \cO^{\g}_{ \a , \b ; i,  j } &= 
C_2 ( \alpha ) ~ \cO^{\g}_{ \a , \b ; i,  j } \cr 
 tr ( Ad_{ G_3}^2) ~ \cO^{\g}_{ \a , \b ; i,  j }     &=~~~~~~~~~~ 
 Ad_{ (G_3)^i_j} Ad_{ (G_3)^j_i}    ~ \cO^{\g}_{ \a , \b ; i,  j } &= 
C_2 ( \beta )~ \cO^{\g}_{ \a , \b ; i,  j }  \cr 
 tr (Ad_{ G_2}^2 Ad_{  G_3})  ~ \cO^{\g}_{ \a , \b ; i,  j }   &=
   Ad_{ (G_2)^i_j}  Ad_{ (G_2)^j_k} Ad_{ (G_3)^k_i} ~
  \cO^{\g}_{ \a , \b ; i,  j }  &= 
C_{223} (\g ;  \a , \b ,  i )~ \cO^{\g}_{ \a , \b ; i,  j } \cr
  tr( Ad_{ G_1}^2 Ad_{  G_4 })  ~ \cO^{\g}_{ \a , \b ; i,  j } &= 
   Ad_{ (G_1)^i_j}  Ad_{ (G_1)^j_k} Ad_{ (G_4)^k_i} ~
  \cO^{\g}_{ \a , \b ; i,  j }  &= 
C_{114 } (\g ; \a , \b ,  j ) ~ \cO^{\g}_{ \a , \b ; i,  j }
\eea

The Casimirs $ C_2 ( \g ) $ is a $U(N)$ Casimir for the composite 
representation $ \g $. 
The Casimirs $ C_2 ( \alpha ) $ and $C_2( \beta ) $ are $U(N)$ 
Casimirs for the representations $ \alpha$ with $m$ boxes and $ \beta $ 
with $n$ boxes. The Casimirs $ C_{114 } ( j ) , C_{223} ( i )$ 
are less familiar. They can be related to elements of the Brauer algebra
which are invariant under the $ \C ( S_m \times S_n )$  subalgebra, 
and can be used to distinguish different copies of the same representation 
$ ( \alpha , \beta )$ of  $  \C ( S_m \times S_n )$ appearing in the 
reduction of the representation $ \gamma $ of $ B_{ N } ( m , n ) $. 
Examples of these eigenvalues will be computed in section \ref{sec:examples} . 

Note that while $ Ad_{G_B} $ and the Casimir 
$ tr  ( Ad_{ G_B} Ad_{ G_B} ) $ leaves the action invariant, the action 
of $ Ad_{ G_1 } $ does not leave the action invariant. 
It is nevertheless useful in organising the  operators 
at zero coupling, since it is  a symmetry of the 
classical scaling operator. It is also worth noting that
$ tr Ad_{ G_1} Ad_{ G_1}$ has the  rather simple effect
of scaling the  action.

\subsection{Enhanced symmetry and Casimirs  : Restricted Schur Basis for 
$ X , X^{\dagger} $ }
\label{ehsrestschu} 

In \cite{bcd}, 
another complete set of gauge invariant operators 
constructed from $X$ and $X^{\dagger}$ was proposed, where 
the symmetric group was used instead of the Brauer algebra\footnote{
A map, called $ \Sigma $, from 
the Brauer algebra $B_{N}(m,n)$ to 
the group algebra of the symmetric group $S_{m+n}$
was given  in \cite{kr} and several useful algebraic properties described, 
and this map was exploited 
in the construction of the basis 
in \cite{bcd}. 
This map also allows a new holomorphic interpretation  of 
the large $N$ expansion of 2DYM \cite{Kimura:2008gs}. 
}. 
Here we have a basis $  
\cO^{ R }_{ R_1 , R_2 ; i , j } \equiv  tr_{ m, n }(  Q^R_{ R_1, R_2 ; i j } (  \bX \otimes \bX^{\dagger} ))$. 
$R$ is an irreducible representation of the symmetric group $S_{m+n}$, 
and $R_{1}$ and $R_{2}$ are irreducible representations of 
$S_{m}$ and $S_{n}$. The indices $i,j$ here each run over the multiplicity 
of the irrep $ R_1 \otimes R_2$ of $ \C ( S_m \times S_n ) $ 
in the irrep $R$ of $ \C ( S_{ m+n} ) $.  
More details on the basis in Appendix \ref{sec:SymmetricBranchingOperator}. 

Here look at $ G_{r} \equiv  G_1 + G_2$. 
On gauge invariant operators  $ G_r = - G_3 - G_4 $. 

We will show 
\bea\label{casactsrest}  
  tr  ( Ad_{ G_r}^2 )~ \cO^{ R }_{ R_1 , R_2 ; i , j } &= \qquad ~~~ 
Ad_{ (G_r)^i_j }  Ad_{ (G_r)^j_i } ~ \cO^{ R }_{ R_1 , R_2 ; i , j }
  & =  C_2 ( R  ) \cO^{ R }_{ R_1 , R_2 ; i , j } \cr 
  tr  ( Ad_{ G_1 }^2 )~ \cO^{ R }_{ R_1 , R_2 ; i , j } & = \qquad ~~~
Ad_{ (G_1)^i_j }  Ad_{ (G_1)^j_i } ~ \cO^{ R }_{ R_1 , R_2 ; i , j } &= 
   C_2 ( R_2  ) \cO^{ R }_{ R_1 , R_2 ; i , j } \cr 
  tr  ( Ad_{ G_2 }^2 )~ \cO^{ R }_{ R_1 , R_2 ; i , j } & = \qquad  ~~~
Ad_{ (G_2)^i_j }  Ad_{ (G_2)^j_i } ~ \cO^{ R }_{ R_1 , R_2 ; i , j } &= 
   C_2 ( R_1  ) \cO^{ R }_{ R_1 , R_2 ; i , j } \cr 
  tr (Ad_{ G_1}^2  Ad_{  G_2} )~ \cO^{ R }_{ R_1 , R_2 ; i , j } & = 
  Ad_{ (G_1)^i_j}  Ad_{ (G_1)^j_k} Ad_{ (G_2)^k_i} ~ \cO^{ R }_{ R_1 , R_2 ; i , j } 
& = C_{ 112} (R ;  R_1 , R_2 , i ) ~ \cO^{ R }_{ R_1 , R_2 ; i , j } \cr 
  tr (Ad_{ G_3}^2 Ad_{  G_4}) ~ \cO^{ R }_{ R_1 , R_2 ; i , j } & =  Ad_{ (G_3)^i_j} 
 Ad_{ (G_3)^j_k} Ad_{ (G_4)^k_i }  ~ \cO^{ R }_{ R_1 , R_2 ; i , j } & =
 C_{ 334} (R; R_1, R_2 ,  j ) ~
 \cO^{ R }_{ R_1 , R_2 ; i , j }  \nn \\
\eea 

$C_2(R)$ is a Casimir of $U(N)$ with $m+n$ boxes.
$C_2(R_1) $ is a Casimir of $U(N)$ for $R_1$ with $m$ boxes. 
$C_2(R_2) $ is a Casimir of $U(N)$ for $R_2$ with $n$ boxes. 
$C_{112} ( R ; R_1 , R_2 , i )  $ and  $C_{ 334 } ( R ; R_1 , R_2 , j )  $
are less familiar. They can be related to elements in $ \C  ( S_{m+n} ) $ 
invariant under $ \C ( S_m )  \times \C ( S_n ) $. They can distinguish 
between different copies of representations $ ( R_1 , R_2 ) $ of 
the product group in the reduction of a fixed $R$. Examples of these 
eigenvalues will be computed in section \ref{sec:examples} . 

From the point of view of the Casimirs it is clear why the 
restricted Schur basis is different from the Brauer basis. 
The Casimir $ tr  ( Ad_{ G_2} Ad_{G_3} )  $ appears in 
the expansion of $ tr ( Ad_{G_B} Ad_{ G_B})  $ which measures 
$ \gamma $ of the Brauer basis. The Casimir $tr ( Ad_{G_1} Ad_{ G_2} )  $
appears in the expansion of  $ tr ( Ad_{G_r} Ad_{ G_r} ) $ 
which measures the $R$-label of the restricted Schur basis. 
It is easy to check that 
\bea 
[  tr ( Ad_{ G_1} Ad_{ G_2} ) ,tr ( Ad_{ G_2} Ad_{ G_3} )  ]  
= tr ( Ad_{ G_1} Ad_{ G_2} Ad_{ G_3 } ) -
   tr ( Ad_{ G_1} Ad_{ G_3} Ad_{ G_2 } ) 
\eea 
Since the two sets of Casimirs do not commute, their 
eigenstates are different.  

%%%%%%%%%%%%%%%%%%%%%%%%%%%%

\subsection{Enhanced symmetries and Casimirs  : $U(M )\times S_n $ basis for 
holomorphic functions of   complex matrices   }\label{sec:umsnI}  

Now we are looking at $M$  complex matrices $ X_1 , X_2 \cdots X_M  $. 
Let us focus on the holomorphic sector. 
We can consider two  different diagonalisations. 
One based on \cite{bhr} which keeps $U(M)$ manifest. 
One based on \cite{bcd} which keeps additional $ U(N)$ symmetries 
manifest. The discussion of the Casimir-diagonal basis 
connection for the restricted Schur basis proceeds by a straightforward 
generalisation of section \ref{ehsrestschu}. We will focus our attention 
on  the basis which keeps $U(M)$ manifest  \cite{bhr}. 

The operators are 
\bea\label{theuMops}  
\cO^{ \Lambda, R , \tau , \beta , \mu } 
& =& \frac{1}{n!}
\sum_{ \alpha } B_{k \beta }  ~ D^R_{ij} ( \alpha )
  ~    C^{ \tau  , \Lambda, k  }_{ R , R , i , j } ~
  tr ( \alpha \bf X ) \cr 
& =& B_{k \beta } C^{ \tau  , \Lambda, k  }_{ R , R , i , j } ~ 
 tr ( Q^R_{ij} { \bf X}  ) 
\eea 
where we defined 
\bea 
 Q^R_{ i j } = \frac{1}{n!}\sum_{ \alpha } D^R_{ij} ( \alpha ) ~ \alpha 
\eea 
and $ { \bf X }  $ is defined as $ X_1 \otimes \cdots \otimes X_1 \otimes X_2 \otimes \cdots \otimes X_2 \otimes \cdots \otimes X_M \otimes \cdots \otimes X_M $ 
with $ \mu_1 $ copies of $ X_1$, $ \mu_2$ copies of $ X_2$ etc. 
and  $ \mu_1 + \mu_2 + \cdots + \mu_M = n $. Here the indices 
$i,j$ are summed  over states in the irrep $R$ of $S_n$, the index 
$k$ is summed over states in the irrep $\L$ of $S_n$. The label $\tau$ 
runs over the multiplicity of $\Lambda$ in the tensor product $R \otimes R $. 
The numbers  $ C^{ \tau  , \Lambda, k  }_{ R , R , i , j } $ are 
Clebsch-Gordan coefficients. The coefficients $ B_{k \beta } $
 are branching coefficients.

The label $ \L$ 
belongs to $U(M)$ or by SW duality to $S_n$. 
 There are generators 
$G_{ab} = \int ~ tr ( X_a \Pi_{X_b} )  $ 
which transform the $a$ indices of $ (X_{a})^i_j$. 
Quadratic Casimirs measuring $ \L$ are $Ad_{ G_{ab} } Ad_{ G_{ba} } $. 
The content $ \mu $  is measured by the Cartan elements $ G_{aa} $. 
The $ R$ label is related to Casimirs of 
$ G ( \cL , X_1  ) + G ( \cL , X_2  ) + \cdots  + 
 G ( \cL , X_M   ) \equiv G_{\cL}  $. Similarly we have $ G_{\cR}  $  
for the right action which also gives the same Casimir. 
This is essentially 
because we can think of the $Q^R_{ij } $ above as $ |R, i \ra \la R , j | $.   
The $ \beta $ label has to do with the reduction from 
$ U ( M ) \rightarrow  U ( 1)^M $  or
equivalently by SW duality  $ S_n \rightarrow S_{\mu_1} \times S_{ \mu_2 }\times \cdots \times S_{\mu_M} $. 
In the special case  
$ M=2$, we denote $ X_1 , X_2 $ as $ X , Y $. The last equation 
below is pretty clear from our discussions of the restricted Schurs, 
 where a similar symmetric group reduction multiplicity appears.

We want to show that we can choose bases $ \tau , \beta $ which satisfy 
\bea\label{toshow}  
  tr ( Ad_{ G_{ \cL }  }   Ad_{ G_{ \cL } }  )  ~ \cO^{ \Lambda, R , \tau , \beta , \mu }  & \equiv &   Ad_{  ( G_{ \cL } ){}^i_j } Ad_{  ( G_{ \cL } ){}^j_i  } 
      ~ \cO^{ \Lambda, R , \tau , \beta , \mu } \cr 
& = &  C_2 ( R ) ~ \cO^{ \Lambda, R , \tau , \beta , \mu } \cr 
 tr ( Ad_{ G_{ \cR }  }   Ad_{ G_{ \cR } }  ) 
 ~ \cO^{ \Lambda, R , \tau , \beta , \mu } 
&\equiv &  Ad_{  ( G_{ \cR } ){}^i_j } Ad_{  ( G_{ \cR } ){}^j_i  } 
  ~ \cO^{ \Lambda, R , \tau , \beta , \mu } \cr 
& = &  C_2 ( R ) ~ \cO^{ \Lambda, R , \tau , \beta , \mu }  
 \cr 
  tr \otimes tr    ( Ad_{ G_{ \cR  } }  Ad_{  E } Ad_{ G_{ \cL } } )   ~ 
 \cO^{ \Lambda, R , \tau , \beta , \mu } & 
\equiv &   Ad_{  ( G_{ \cR } ){}^{l}_j } Ad_{ E^{jk}_{lm} }
 Ad_{  ( G_{ \cR } ){}^{m}_k }   ~  \cO^{ \Lambda, R , \tau , \beta , \mu } \cr 
&=&  C_{T} ( \tau )   ~ \cO^{ \Lambda, R , \tau , \beta , \mu } \cr 
 tr (  Ad_{  G (\cL , X ) }   Ad_{G ( \cL , X ) } Ad_{  G ( \cL , Y ) }   )
~ \cO^{ \Lambda, R , \tau , \beta , \mu } 
 &\equiv & Ad_{  (G (\cL , X ))^i_j }   Ad_{( G ( \cL , X ))^j_k } Ad_{  (G ( \cL , Y ))^k_i }   
~ \cO^{ \Lambda, R , \tau , \beta , \mu } \cr 
&=&  C_{XXY} ( \beta ) ~  \cO^{ \Lambda, R , \tau , \beta , \mu }
\eea

\subsection{Diagonal bases and Casimirs : the general case of
              $ G \times S_n$ }\label{sec:generalcase}

A general construction of diagonal bases of operators 
compatible with global symmetry group $ G $ using Clebsch-Gordan coefficients 
of $ G \times S_n $ was given \cite{bhr2}. 
The operators are of the form 
$ \cO^{ \Lambda, \Lambda_1 ,  R , M_{\Lambda} , \tau_{\Lambda_1 \Lambda} , \tau  }$. 
 In a theory with symmetry group $G$ there 
will be corresponding 
charges $ Q_a $ acting as $Ad_{ Q_a} $  via commutators. 
Casimirs such as $ Ad_{ Q_a } Ad_{ Q_a  } $, where the indices are 
contracted with a Killing form on the Lie algebra, give 
quadratic Casimirs which measure $\Lambda$, which labels irreps. 
of $G$. 
 Higher order invariants  give 
higher Casimirs sensitive to $ \Lambda$. The $Q_a$ will include 
a maximally commuting Cartan sub-algebra which measure the states
 $ M_{ \Lambda} $ in the irrep. $ \Lambda$. 
 The $S_n$ representation $ R  $ corresponds to a Young 
diagram with $n$ boxes. As in earlier sections 
there are Casimirs constructed from the $U(N)$ generators acting on 
the upper  (or lower) indices  of the matrix fields, which measure $R$. 
The $ \Lambda_{ 1} $ 
label corresponds to an $S_n$ which acts simultaneously on the 
upper and lower $U(N)$ indices of the matrix fields. Hence the 
Casimirs constructed  from this diagonal  $U(N)$ measure  $\Lambda_1$.
The label $ \tau $ runs over the multiplicity of $ \Lambda_1 $ in the 
Clebsch-Gordan decomposition of the tensor product $ R \otimes R $. 
The invariants for can be obtained from $U(N^2)$ and $U(N) \times U(N)$ 
generators, as in sections \ref{sec:umsnI} and \ref{sec:umsnII}. 
Finally there is the label $ \tau_{\Lambda_1 \Lambda}$ which runs over the 
multiplicity of the $ G \times S_n$ representation 
$ V_{ \Lambda} \otimes V_{ \Lambda_1 } $ in the decomposition of 
$ V_F^{ \otimes n } $, where $V_F$ is the representation of $G$  carried
 by the fundamental fields. We expect that similar ideas can be used 
to construct Casimirs which distinguish this multiplicity. To make this 
more precise we need a better understanding of the commutant 
$ Com ( G \times S_n ) $ of 
$ G \times S_n$ in $ V_F^{ \otimes n } $. A Cartan-like subalgebra will 
distinguish states in the multiplicity space $ V_{ \Lambda, \Lambda_1} $ which 
form a representation of  $ Com ( G \times S_n ) $. For the symmetric group 
a Cartan-like sub-algebra plays a prominent role in the Vershik-Okunkov 
approach \cite{VershikOkounkov} to symmetric group
 representations. We will use this approach in 
section \ref{sec:examples}. A generalisation of these Cartan-like 
sub-algebras to  commutant algebras such as $ Com (  G \times S_n ) $ 
and their  expression in terms of Noether charges constructed from
iterated applications of  the generators of $G$ would lead to
the resolution  of  the parameter $ \tau_{ \Lambda,  \Lambda_1 } $. 
In other words it would lead to Casimir-like operators which 
have the $ \cO^{ \Lambda, \Lambda_1 ,  R , M_{\Lambda} , \tau_{\Lambda_1 \Lambda} , \tau  }$
as eigenvectors and whose eigenvalues will depend non-trivially 
on $ \tau_{\Lambda_1 \Lambda}$. In the case of the full field content of 
$\cN=4 $ SYM we would want to 
implement the above programme for $ G = PSU(2,2|4)$. It seems likely that 
this would show up some interesting connections  to Yangians and 
other approaches to integrability in AdS/CFT  along the lines 
of \cite{dolnapwit}.

\section{Diagonal bases of gauge invariant operators  as eigenstates of Casimirs : Proofs  } 

\subsection{Brauer Basis :  One  complex matrix }\label{sec:PfsBrauerXX*}   

We prove the claims in section \ref{sec:ehsBrauer}
that the labels on the diagonal basis of operators 
$ \cO^{\gamma}_{ \alpha , \beta , i , j }  $ are 
measured by Casimirs of enhanced symmetries. 

\subsubsection{The $U(N)$ generators  and $ C_2 ( \g ) $ }
\label{PfsBrauerXXgamm}  

We will prove the first claim in section \ref{sec:ehsBrauer} 
that Casimirs constructed from  $ G_B = G_2 + G_3 $ 
measure the label $ \gamma $. 
 The action of $G_B$ is 
\bea 
&& [ (G_B){}^i_j , X^p_q ] = \delta_j^p X^i_q \cr 
&& [  ( G_B){}^i_j , ( X^*)^p_q ] = - \delta^{ip}  (X^*)^j_q 
\eea 
From (\ref{UNcomrels}), we see that  $ { G_B}^i_j $ indeed obey 
relations of $U(N)$ 
\bea 
[ (G_B){}^i_j , (G_B){}^k_l ] 
= \delta^k_j (G_B){}^i_l - \delta^i_l (G_B){}^k_j 
\label{relationUN}
\eea 
Defining the field theory commutator  
 $Ad_{ (G_B){}^i_j }  \equiv [(G_B){}^i_j,\cdot]$, we see that 
 $Ad_{ (G_B){}^i_j} Ad_{ (G_B){}^j_i }  \equiv \hat{{\bold C}}_2 $
 acts on the upper indices of $ X \otimes X^*$ 
the way  the quadratic Casimir $\hat C_2 \equiv E^i_j E^j_i $
 of $ U(N)$ acts on $ V \otimes \bar V $.

We also have the charges $ \tilde G_B = G_1 + G_4$
which act on the fields as 
\begin{eqnarray}
&& [ (\tilde G_B ) {}^i_j , X^p_q ] = -  \delta^i_q X^p_j \cr 
&& [ (\tilde G_B ) {}^i_j , ( X^*)^p_q ] =  \delta_{jq}( X^*)^p_i
\end{eqnarray}
The  charge $\tilde G_{B} $ also satisfies the same commutation relation as 
(\ref{relationUN}). 
As long as 
we act with $(G_B)$ and $(\tilde G_B ) $ on gauge invariant operators,  
we have $Ad_{ (G_B) } + Ad_{ (\tilde G_B )  } =0  $
 because $Ad_{ (G_B) }  + Ad_{ (\tilde G_B )  } $ 
generates the global adjoint gauge transformation 
$X \rightarrow UX U^{\dagger}$, 
$X^{\dagger} \rightarrow UX^{\dagger} U^{\dagger}$. 

The action of $ { \bf \hat C}_2 \equiv  Ad_{ G_B } Ad_{ G_B} $ on  
$ X^{p_1}_{q_1}  \cdots   X^{p_m }_{q_m}   
 {X^*}^{p_{m+1} }_{q_{m+1}} 
 \cdots {X^*}^{p_{m+n }}_{q_{m+n}  } $  leaves fixed the 
lower indices and acts on the upper indices just the way 
$\hat{C}_2$ acts on $V^{\otimes m} \otimes {\bar V }^{\otimes  n }   $. 
Now we know, from Schur-Weyl duality that 
$V^{\otimes m} \otimes {\bar V }^{\otimes  n }   $
decomposes into 
\bea 
V^{\otimes m} \otimes {\bar V }^{\otimes  n }   
= \bigoplus_{\gamma} V_{\gamma}^{ U(N)}  \otimes V_{\gamma}^{ B_{N}(m,n) } 
\eea 
The irrep. $ V_{\gamma}^{B(m,n) }  $ , which gives the 
multiplicity space of $V_{\gamma}^{ U(N) } $, decomposes 
into a number of irreps. of $ \mathbb C (S_{m}\times S_{n})$. 
The branching operators leave the $V_{\gamma}^{ U(N) } $ 
factor unchanged, and map from one irrep of the  
$ \mathbb C (S_{m}\times S_{n})$ 
to another. Hence we have
\begin{eqnarray}
 Q^{ \gamma}_{ A,   ij  }\left(V^{m}\otimes \bar{V}^n \right)
=V_{\gamma}^{U(N)}\otimes 
Q^{ \gamma}_{A,  i j  }\left(V_{\gamma}^{B_{N}(m,n)}\right)
\end{eqnarray}
This means that the action of the $U(N) $ generators $ ( G_B) $  on  
$ Q^{\gamma}_{A, ij}  ( V^m  \otimes \bar V^n )  $ is the same as 
on the irrep $ \gamma $ of $U(N)$. Hence it follows 
that 
\bea 
{ \bf \hat{C}}_2  ~ Q^{ \gamma}_{A,ij }  ( \bold{X} \otimes \bold{X}^* )  
 = C_2 ( \gamma )  ~  Q^{ \gamma }_{A ,ij  } ( \bold{X} \otimes \bold{X}^* ) 
\eea 
and we obtain the first line of (\ref{bbascaseqs}).

There are higher order Casimirs such as 
$ { \bf  \hat C }_{3}  \equiv 
 Ad_{ (G_B)^i_j }  Ad_{ (G_B)^j_k }   Ad_{ (G_B)^k_i } $ 
\begin{eqnarray}
&&
{ \bf \hat{C}}_{3}  ~
tr_{m,n}\left(Q^{\gamma}_{A,ij}
(\bold{X}\otimes \bold{X}^{\ast})
\right)
= C_{3}(\gamma) ~ 
tr_{m,n}\left(Q^{\gamma}_{A,ij}
(\bold{X} \otimes \bold{X}^{\ast})
\right)
\end{eqnarray}
It is easy to check directly in simple cases, e.g $m=n=1$,
that the above equations holds. 
Consider  the case $m = 1, n = 1$ 
\bea 
&& \hat{\bold{C}}_2  ~ tr X X^{\dagger}   = 0 \cr 
&& \hat{\bold{C}}_2  ~ tr X tr X^{\dagger}   = 
2N \left(  tr X tr X^{\dagger} -
 { 1 \over N }   tr X X^{\dagger}  \right)
\eea
Hence the eigenstates of $\hat{\bold{C}}_2$  are  $ tr X X^{\dagger} $ 
and $ (tr X tr X^{\dagger} - { 1 \over N }  tr X X^{\dagger} ) $, 
which are the simplest examples of operators 
$ \cO^{ \g}_{ \a , \b ; i , j } $, 
see (A.37) in \cite{kr}.

%%%%%%%%%%%%%%%%%%%%%%%%%%%%%%%%%%%%%%%%%%%%%%%%%%%%%%%%%%%%%%%%%%

\subsubsection{Casimirs  $ C_2( \alpha ) $  and $ C_2 ( \beta ) $ } 
\label{PfsbraXXalphbeta} 

Recall that $ G_B $ is a sum  
\begin{eqnarray}
G_B = G_2 + G_3 
\end{eqnarray}
The actions on the fields are as follows
\begin{eqnarray}
&& [ ( G_2)^i_j   , X^p_q ] = \delta_j^p X^i_q \cr 
&& [  ( G_2)^i_j  , ( X^*)^p_q ] =0 \cr
&& [ ( G_3)^i_j   , X^p_q ] = 0\cr 
&& [  ( G_3)^i_j  , ( X^*)^p_q ] =-\delta^{ip}X^{j}_{q} 
\end{eqnarray}
We can easily check that $(G_2){}^i_j$ and $(G_3){}^i_j$
also obey the commutation relation of $U(N)$.
 Therefore we also have a Casimir which is given by 
\begin{eqnarray}
{ \bf \hat{C}}_{2}^{++}  \equiv Ad_{ (G_2){}^i_j } Ad_{ (G_2){}^j_i } 
\end{eqnarray}
Similarly we have another Casimir 
\begin{eqnarray}
{ \bf \hat{C}}_{2}^{--}   \equiv Ad_{ ( G_3 ){}^i_j }  Ad_{ (G_3){}^j_i }
\end{eqnarray}

We will prove 
\begin{eqnarray}
&&
{\bf \hat{C}}_{2}^{++} ~
tr\left(Q^{\gamma}_{(\alpha,\beta),ij}
(\bX\otimes \bX^{\ast})
\right)
= C_{2}(\alpha) tr\left(Q^{\gamma}_{(\alpha,\beta),ij}
(\bX\otimes \bX^{\ast})
\right)
\cr
&&
{\bf \hat{C}}_{2}^{--}~
tr\left(Q^{\gamma}_{(\alpha,\beta),ij}
(\bX\otimes \bX^{\ast})
\right)
=C_{2}(\beta)
tr\left(Q^{\gamma}_{(\alpha,\beta),ij}
(\bX\otimes \bX^{\ast})
\right)
\end{eqnarray}

The following steps give the result
\bea 
&&
 {\bf  \hat C_2 }^{++}   ( Q^{ \gamma }_{ (\alpha,\beta) , i j } )^I_{ J }
 ( \bX \otimes \bX^* )^{J}_K 
\cr && =  ( { \hat C_{2}}^{++}  )_L^J ( Q^{ \gamma }_{ (\alpha,\beta) , i  j } )^I_{ J }
( \bX \otimes \bX^* )^{L}_K
\cr && = (  \hat C_{2}^{++}  Q^{ \gamma }_{ (\alpha,\beta) , i  j } )^{ I}_L
 ( \bX \otimes \bX^* )^{L}_K
\cr && =  C_{2} ( \alpha )  ( Q^{ \gamma }_{ (\alpha,\beta) , i  j } )^{ I}_L
  ( \bX \otimes \bX^* )^{L}_K
\eea 
where $\hat C_{2}^{++}=\sum_{r,s=1}^{m}\rho_{s}(E^{i}_{j})
\rho_{t}(E^{j}_{i})$ and 
$\hat C_{2}^{--}=\sum_{r,s=m+1}^{m+n}\rho_{s}(E^{i}_{j})
\rho_{t}(E^{j}_{i})$.

To get the last equality we can use the expression 
$  Q^{ \gamma }_{ A , i j } = 
\sum_{m_{A}}
|\gamma \rightarrow A , m_A ,  i \ra \la\gamma \rightarrow A , m_A ,  j|  $
or the expression in terms of the algebra
\bea 
Q^{\gamma}_{A , i j } = \sum_{ b}   B_{ \g  , A , I , i }^{  \dagger}  
  D^{\gamma}_{ IJ} ( b ) B_{ \gamma , A , J , j }  b^*   
\eea 
This expression is derived in the Appendix \ref{sec:SymmetricBranchingOperator}.

%%%%%%%%%%%%%%%%%%%%%%%%%%%%%%%%%%%%%%%%%%%%%%%%%%%%%%%%%%%%%%%

\subsubsection{$\hat { \bf{C}}_{223 }$ and $ \hat {  \bold{C}}_{114} $   : Casimir for $ i , j $  labels of 
$ Q^{ \gamma}_{ \alpha , \beta ; i j } $   } 
\label{PfsbraXXij}

The commutator action of the generators $G_2$ and $G_3$ act on the upper index 
of $X $ and $ X^* $ while leaving the lower indices unchanged 
(see (\ref{epsact}) and (\ref{unact})). 
When we have a sequence of $m$ copies of $X$'s and $n$ copies of $X^*$, 
the set of upper indices transforms as 
$ V^{ \otimes m } \otimes \bar V^{ \otimes n } $, were $V$ is the fundamental 
and $ \bar V $ the anti-fundamental of $U(N)$. The action 
of $ { \bf \hat C}_{223 } \equiv Ad_{ (G_2)^i_j}  Ad_{ (G_2)^j_k} Ad_{ (G_3)^k_i} $  
is the same as the action of 
$ \rho_{\underline{m}}  (E^{i}_{j} ) 
\rho_{\underline{m}}  ( E^{j}_{k}  )
 \rho_{\underline{n}}  (E^{k}_{i}  ) $, 
 where $\rho_{\underline{m}}   ( E ) $ denotes 
the action of $U(N)$ generators on $V^{\otimes m } $ and 
$ \rho_{\underline{n}} (E) $ denotes the action of $U(N)$ 
generators on $\bar V^{\otimes n } $. 

\begin{eqnarray}\label{cas++-} 
\hat{C}_{++-}&=&
\sum_{r,s=1}^{m}\sum_{t=m+1}^{n}
\rho_{r}(E^{i}_{j})
\rho_{s}(E^{j}_{k})
\rho_{t}(E^{k}_{i}) \cr
%&=&\sum_{r,t}
%\rho_{r}(E^{+}{}^{i}_{j}E^{+}{}^{j}_{k})
%\rho_{t}(E^{-}{}^{k}_{i})
%+\sum_{r\neq s,t}
%\rho_{r}(E^{+}{}^{i}_{j})
%\rho_{s}(E^{+}{}^{j}_{k})
%\rho_{t}(E^{-}{}^{k}_{i})  \cr
%&=&-N\sum_{r,t}C_{r\bar{t}}
%-\sum_{r\neq s,t}(rs)C_{r\bar{t}} \cr
&=&-N\sum_{r,t}\left(
1+\frac{1}{N}\sum_{s\neq r}(rs)
\right)C_{r\bar{t}}
\label{c++-Brauer}
\end{eqnarray}
This  will be shown in the appendix \ref{CasimirBrauer}.

The action ${ \bf \hat C}_{223 }$ on the operators is  
\bea\label{pf223}  
&& { \bf \hat C}_{223 }   ( Q^{\gamma}_{ A , i j } )^J_I  ( \bX \otimes \bX^* )^I_J  
\cr 
&& =  ( Q^{\gamma}_{ A , i j } )^J_I  ( \hat C_{ ++- }  )^{ I}_K
  ( \bX \otimes \bX^* )^K_J 
\cr 
&& = ( \hat C_{ ++- } Q^{\gamma}_{ A , i j } )^J_I  ( \bX \otimes \bX^* )^K_J \cr 
&& = ( \hat C_{  ++- } | \gamma , A , m_A , i \ra \la  \gamma , A , m_A , j | )^{J}_I 
 ( \bX \otimes \bX^* )^K_J \cr 
&&  = C_{ ++- } ( \gamma , A , i )  ( Q^{\gamma}_{ A , i j } )^J_I
( \bX \otimes \bX^* )^K_J  
\eea

Note that $ Ad_{ G_1}  $ and $Ad_{ G_4} $ act on the lower indices of $ X , X^*$. 
To measure $j$ we consider 
\bea 
&& { \bf \hat C_{114 } }  ( Q^{\gamma}_{ A , i j } )^J_I  ( \bX \otimes \bX^* )^I_J  
\cr 
&& =  ( Q^{\gamma}_{ A , i j } )^J_I  ( \hat C_{ ++- }  )^{K  }_J
  ( \bX \otimes \bX^* )^I_K 
\cr 
&& = ( Q^{\gamma}_{ A , i j } \hat C_{ ++- }  )^K_I ( \bX \otimes \bX^* )^I_K \cr 
&& = C_{++- } ( \g, A , j ) ( Q^{\gamma}_{ A , i j } )^K_I ( \bX \otimes \bX^* )^I_K
\eea

In section \ref{sec:examples}, we show explicitly how to find the basis 
of states which diagonalise the Casimir $ \hat C_{ ++-} $
and the eigenvalues  $  C_{ ++-} ( \gamma , A , i ) $ in a simple 
 example.

%%%%%%%%%%%%%%%%%%%%%%%%%%%%%%%%%%%%%%%%%%%%%%%%%%%%%%%%

\subsection{Casimirs and restricted Schur Basis  for $ X , X^{\dagger} $  
    }\label{PfssymmXX*}  
 
Instead of using the Brauer algebra $ B_N( m , n ) $ and its 
reduction to the sub-algebra $ \C (  S_m )  \times \C (  S_n )  $
we are now using the algebra $ \C ( S_{ n + m } ) $ and its 
reduction to  $ \C (  S_m )  \times \C (  S_n )  $. The Brauer 
algebra is the commutant of $ U(N) $ acting on 
$ V^{ \otimes m } \otimes \bar V^{\otimes n  } $.
 This representation is 
equivalent to the action of $G_2, G_3 $ (see (\ref{unact}))
 on the upper indices of 
$ X , X^{ \ast } $. The lower indices are similarly acted on by $G_1 , G_3 $. 
On the other hand,   $ \C ( S_{ m + n } ) $ 
is dual to the action of $ U(N)$ on $ V^{ \otimes ( m +n )  } $. 
This is equivalent to the action of $ G_1 , G_2 $  (see (\ref{unact}))
on the upper indices of $ X , X^{\dagger} $.  
The lower indices are similarly acted on by $ G_3 , G_4 $. 
The space of gauge invariant operators can be constructed as 
$ tr_{ m,n} ( b ( \bX \otimes \bX^* ) )  $ or as 
$tr_{ m+n }  ( \sigma ( \bX \otimes \bX^{\dagger} )) $ :
 the two constructions can be  related via the invertible
 map $ \Sigma $ from Brauer algebra to symmetric group, but give distinct 
 bases diagonalising the two-point functions.   
The calculations are closely parallel, with the role of 
the pair $ ( G_2 , G_3 ) $ of section \ref{sec:PfsBrauerXX*} 
 being now played by the pair $ ( G_1 , G_2 ) $,
 and the pair $ (G_1, G_4 ) $ of section 
 \ref{sec:PfsBrauerXX*}  now played by $ ( G_3 , G_4 )  $.

$ tr  ( Ad_{ G_r } Ad_{ G_r } )  $ acts on the 
upper indices of $m$ copies on $ X $ and $n$ copies of $X^{\dagger } $
as the Casimir of $U(N)$ acting on $ V^{ \otimes ( m +n ) } $. 
This Casimir action can be expressed in terms of 
 symmetric group action (a step which is useful in the string theory 
of two dimensional Yang Mills \cite{cmr})
\begin{eqnarray}
\hat{C}_{2}
&=&\sum_{r,s}\rho_{r}(E^{i}_{j})\rho_{s}(E^{j}_{i})\cr
&=&(m+n)N + \sum_{r\neq s}(rs)
\end{eqnarray}
The derivation of this is in the Appendix \ref{sec:UNtoFinitealgebra}.
This relation between the 
$U(N)$ group and the symmetric group follows essentially from 
Schur-Weyl duality 
\bea 
 V^{ \otimes m + n } = \bigoplus_{R} V_R^{U(N) }  \otimes V_R^{ S_{m+n } }
\eea 
Using the analogous steps to  section \ref{PfsBrauerXXgamm} we see 
\bea 
&&
 { \bf \hat C}_2  ( G_r  )  ( Q^{R}_{ R_1, R_2 ; ij } )^I_J    
( \bX \otimes \bX^{\dagger} )^J_I  
= C_2 ( R )   ( Q^{R}_{ R_1, R_2 ; ij } )^I_J  
( \bX \otimes \bX^{\dagger} )^J_I 
\eea 
hence proving the first line of (\ref{casactsrest})

Now we turn to the measurement of the quantum numbers $R_1 , R_2 $ 
in the restricted Schur basis. The operator 
$ tr ( Ad_{G_1 } Ad_{ G_1 } ) $ acts like the Casimir 
$ \rho_{ \underline{ n } }  ( E^i_j   ) \rho_{ \underline{ n}  }  ( E^j_i  )  $. 
The  
$\rho_{ \underline{n} } (  E{}^{i}_{j} )$ denotes the action of $U(N)$ on the 
$n$ fold tensor product of the fundamental. 
This can be expressed as $ Nn + \sum_{ r \ne s = 1 }^n (rs) $ 
in the group algebra of $S_n$. As in section \ref{PfsbraXXalphbeta} 
the action of the $ \C (S_n ) $ element on $ Q^{R}_{ R_1 , R_2 ; i , j } $ 
results in $ C_2 ( R_2 ) $. Likewise   $ tr ( Ad_{G_2 } Ad_{ G_2 } ) $
gives $ C_2 ( R_1 )  $.

The operator $ tr  Ad_{ G_1 } Ad_{ G_1 } Ad_{ G_2 } $ 
acts  on the upper indices of $ X^{i_1}_{ j_1} \cdots X^{i_m}_{j_m} 
  X^{ \dagger }{}^{i_{m+1}  }_{j_{m+1}} \cdots X^{\dagger}{}^{i_{m+n} }_{ j_{m+n} }  $
as $\rho_{ \underline{n} } (  E{}^{i}_{j} ) \rho_{ \underline{n}  } 
(  E{}^{j}_{k} ) 
\rho_{ \underline{m}  } (  E{}^{k}_{i} ) \equiv \hat C_{ 112 }  $ acting on 
$ V^{ \otimes m } \otimes V^{ \otimes n } $. 
\begin{eqnarray}
{ \hat{C}_{112} } &=&
\sum_{r=m+1}^{m+n} \sum_{s=m+1}^{m+n}  \sum_{t=1}^{m} 
\rho_{r }(E{}^{i}_{j})
\rho_{s}(E{}^{j}_{k})
\rho_{t}(E{}^{k}_{i}) \cr
&=&N\sum_{r,t}\left(1+\frac{1}{N}\sum_{s\neq r}(rs)\right)(rt)
\end{eqnarray}
The element 
$ \sum_{r=m+1}^{m+n} \sum_{s=m+1}^{m+n}  \sum_{t=1}^{m} ( rs ) ( rt  ) $ 
is in $ \C ( S_{ m + n } ) $ but is invariant under the subalgebra 
$  \C ( S_{ m } ) \times \C ( S_{ n } )$. It has matrix elements  
\bea 
\la R \rightarrow  R_1 , R_2 ; m'_{R_1} , m'_{R_2} , i'     | \hat C^{ 112}  | R \rightarrow  R_1 , R_2 , m_{ R_1 } , m_{ R_2 } , i \ra 
     =   ( C_{ 112 }  )_{ i' i } \delta_{ m_{R_1 } , m'_{ R_1} }
\delta_{ m_{R_2 } , m'_{ R_2} } 
\label{restrictmultiplicity}
\eea 
By diagonalising the matrix   $  ( C _{112})_{ i i'} $ we can find the
eigenstates and eigenvalues, which distinguish different indices $i$. 
An example is given in section \ref{sec:examples}.

%%%%%%%%%%%%%%%%%%%%%%%%%%%%%%%%%%%%%

\subsection{$ U(M) \times S_n$ basis : for holomorphic 
$ X_1 , X_2 , \cdots , X_M  $ } 
\label{sec:umsnII} 
 
\subsubsection{The $ \tau $ multiplicity and the $U(N^2)$ symmetry }

The $ \tau $ label runs over  the multiplicity of $ \Lambda$ 
in the Clebsch-Gordan decomposition of $ R \otimes  R $.
Another way to say this is that it runs over  
 the multiplicity of the irrep $ \Lambda$ of the diagonal  subgroup 
$S_n$  in the irrep  $ R \otimes R $ of  $ S_n \times S_n $. 
The upper and lower indices of the $X$'s  in equation (\ref{theuMops})
have been projected to $R$ due to the insertion of $Q^R_{ij}$. 
We can think of $ Q^R_{ ij} $ as $ |R, i \ra \la  R , j | $. 
The product group $S_n \times S_n$ is SW dual to $U(N) \times U(N)$. 
where the two $ U(N) $ act on upper and lower indices respectively.
 The action on upper and lower indices is generated by $ G_{\cL} $ and  $ G_{ \cR } $
defined in section \ref{sec:umsnI}. 
   The operators in equation  (\ref{theuMops}) are linear combinations of 
\bea 
 ( X_{a_1} )^{p_1}_{q_1}   ( X_{a_2} ) ^{p_2}_{q_2} \cdots  ( X_{a_n})^{p_n}_{q_n}
\eea
To be more precise we take $ a_1, \cdots ,a_{\mu_1} = 1 $, 
$ a_{\mu_1+1 }, \cdots ,a_{\mu_2 } = 2 $ etc. 
The symmetry generator of interest $ G_{ \cL }$  acts in the same way 
irrespective of what value the $a_i$ have.
The diagonal $S_n $ is SW dual to $U(N^2) $. By the general theorem 
discussed in the introduction, this means that the $ \tau $ multiplicity 
is related to the reduction of $ U(N^2)$ to $ U(N) \times U(N) $.   
The $U(N^2)$ has been discussed in section \ref{sec:UNsq}.

Consider therefore the  operator 
$ Ad_{ E^{jk}_{lm} } Ad_{  ( G_{\cL} )^m_k } Ad_{   ( G_{ \cR  }  ) ^l_j } $ 
on the above. This is made from the $ U(N) \times U(N)$  along with the 
$U(N^2)$ and is invariant under the global gauge symmetry since it 
has all indices contracted. 
 In other words we are considering 
\bea 
[  E^{jk}_{lm} , [  ( G_{\cL} )^m_k , [  ( G_{ \cR  }  ) ^l_j , 
   { \bf X } ] ]]
\eea 

The adjoint action of each symmetry generator  on $ { \bf X }$ 
produces a sum of $n$ terms, each following from the linear transformation 
of different factors in the $n$-fold product. The actions on a single $X$ 
is given by  
\begin{eqnarray}
&&
Ad_{ {E}^{jk}_{lm} } (  X^{p}_{q} ) = 
     [ {E}^{jk}_{lm} ,  X^{p}_{q} ] =\delta^{p}_{m} \delta^{j}_{q}X^{k}_{l} \cr
&&
Ad_{  ( G_{{\cL }} ){}^{l}_{j}    } (  X^{p}_{q} ) =    [  ( G_{{\cL }} ){}^{l}_{j} , X^{p}_{q} ] =\delta^{p}_{j}X^{l}_{q} \cr
&&
A_{ ( G_{{\cR }}) {}^{m}_{k}  }  (  X^{p}_{q} )  =   [ ( G_{{\cR }}) {}^{m}_{k} , X^{p}_{q} ] =\delta^{m}_{q}X^{p}_{k} 
\end{eqnarray}
For any of the symmetry generators 
$ Ad_S ( \bf X  ) $ can be expanded as 
\bea  
Ad_S = \sum_{ i=1}^n   Ad_S^{ (i)} 
\eea 
where   $ Ad_S^{ (i)} $ acts on the $i$'th factor of the $n$-fold product. 

The action of the operator can be expressed in terms 
of elements in the symmetric group
\bea
&&Ad_{   E^{jk}_{lm} } Ad_{   ( G_{\cL} )^m_k } 
Ad_{   ( G_{ \cR  }  ) ^l_j   } 
-NAd_{  ( G_{{\cL}}  ) {}^{m}_{k} } 
A_{  ( G_{{\cL }} ) {}^{k}_{m} }
-NAd_{  ( G_{{\cR}}  ) {}^{l}_{j} } 
A_{  ( G_{{\cR }} ) {}^{j}_{l} }
-Ad_{   E^{jk}_{lm} } 
Ad_{   E^{lm}_{jk} } 
-2N^{2}n
\cr
&=&
\sum_{ r \ne s \ne t } (rs)_{\bu}  (rt)_{\bl}  
 \label{generalisedUN2}
\eea  
The derivation of this is given in the Appendix 
\ref{sec:CasimirUN2}.
We next show that 
the above field theory operator can measure the multiplicity.  

\subsubsection{Proof  } 

In \cite{bhr} the diagonal basis contains $  \tau $ 
which runs over the multiplicity of the irrep. $ \Lambda $ 
of $S_n$ in the tensor product decomposition 
 $ R \otimes R $ of the irrep. $R$ of $S_n$. 
 
The Casimirs $C_{ T} ( \tau ) $ are defined 
generally for tensor products of $ S_n$ representations. 
For  tensor product states $ | R_1  ,R_2 ,  m_1  , m_2 \ra $ 
we have the action of $ \hat C_{ T } \equiv 
 \sum_{ r \ne  s \ne  t } (rs) \otimes  (rt) $ 
as 
\bea\label{CHact}  
\hat C_{ T } | R_1  ,r_2 ,  m_1  , m_2 \ra &=& 
\sum_{ r \ne s \ne t } ( rs) \otimes ( rt ) | R_1  , R_2,  m_1  , m_2 \ra \cr 
& =& \sum_{ r \ne s \ne t } D^{R_1}_{ n_1 m_1 } ( ( rs ) ) 
 D^{R_2}_{ n_2 m_2 } ( ( rt ) )  |  R_1  , R_2 ,n_1 , n_2 \ra
\eea 
The  tensor product space can be decomposed in terms of 
irreps. of $S_n$ using the Clebsch-Gordan coefficients : 
\bea 
| R_1 , R_2 , m_1  , m_2 \ra = C_{ R_1 , R_2 , m_1 , m_2 }^{
\Lambda, m_{ \Lambda} , \tau } | \Lambda, m_{ \Lambda} , \tau \ra 
\eea 
For the application at hand we will be interested in  the 
special case $R_1 = R_2 = R $.
The element  $ \hat C_{ T} $ is an element of  the group algebra 
$ \mathbb{C}  ( S_n ) \times \mathbb{C} ( S_n ) $ which is invariant under the diagonal 
$ \mathbb{ C}  ( S_n ) $.  The Clebsh-Gordan decomposition is equivalent to 
the reduction from the tensor product algebra to its diagonal subalgebra. 
Since $ \hat C_{ T } $ commutes with the diagonal 
$  \mathbb{ C}  ( S_n ) $, 
by Schur's Lemma, its matrix elements in the $S_n$ basis 
satisfy : 
\bea 
\la \Lambda,m_{ \Lambda} , \tau | { \hat C }_{ T} | \Lambda^{\prime} , m'_{\Lambda' } ,
 \tau^{ \prime}   \ra 
= \delta_{ \Lambda\Lambda' } \delta_{ m_{ \Lambda}  m'_{ \Lambda} }
 ( C_{T} )_{  \tau \tau' } 
\label{taugeneralised}
\eea 
By choosing a basis in the multiplicity space such that 
$ ( C_{T} )_{   \tau \tau' } $ 
is diagionalised, we have
\bea 
\la \Lambda,m_{ \Lambda} , \tau | { \hat C }_{ T} | \Lambda^{\prime} , m'_{\Lambda' } ,
 \tau^{ \prime}   \ra 
= \delta_{ \Lambda\Lambda' } \delta_{ m_{ \Lambda}  m'_{ \Lambda} } \delta_{ \tau \tau' } C_{T} ( \tau )  
\eea
Inserting a complete set  of states in the left of 
 (\ref{CHact}) and taking an 
overlap with a state labelled by irreps of the diagonal $ S_n$ : 
\bea\label{CHmatel}  
&&  \la \Lambda, m_{ \Lambda} , \tau |  \hat C_{ T }  | \Lambda' ,
 m'_{ \Lambda' } , \tau^{\prime } \ra \la \Lambda' , m'_{ \Lambda' } , \tau' | R_1 , R_2 , m_1 
, m_2 \ra \cr 
&& =  C_{ T } ( \tau )  
\la \Lambda, m_{ \Lambda} , \tau | R_1 , R_2 , m_1 ,  m_2 \ra  \cr 
&& = C_{R_1 , R_2 ,m_1 ,  m_2 }^ { \Lambda, m_{\Lambda} , \tau }  C_{ T } ( \tau )
\eea 
Taking the same matrix element on the right of  (\ref{CHact})
gives the identity 
\bea\label{tauident} 
 C_{R_1 , R_2 , n_1 , n_2 }^ { \Lambda, m_{\Lambda} , \tau }  C_{ T } ( \tau )
= \sum_{ r \ne s \ne t } D^{R_1}_{ n_1 m_1 } ( ( rs ) ) 
 D^{R_2}_{ n_2 m_2 } ( ( rt ) ) 
C_{ R,  R , m_1 , m_2  }^{ \Lambda, m_{ \Lambda} , \tau }  
\eea 

Now consider the field theory operators. 
In (\ref{generalisedUN2}), 
we expressed the action by commutators of 
certain sequences of enhanced symmetries in terms of generalised  
Casimirs such as the above. We wrote it in terms of 
$ \sum_{ r , s , t } ( rs )_{ \bf { u } }  ( rt )_{ \bf { l } }  $
indicating the action on the upper and lower indices. 
We find 
\bea 
&& \sum_{ r , s , t } ( rs )_{ \bf { u } }  ( rt )_{ \bf { l } } 
  \sum_{ \alpha } B_{k \beta }  ~ D^R_{ij} ( \alpha )
  ~    C^{ \tau  , \Lambda, k  }_{ R , R , i , j } ~   
  ( X_{ a_1} ) {}^{ i_1}_{i_\alpha( 1 ) } \cdots ( X_{a_n}) {}^{i_n}_{i_{\alpha(n)}} 
 \cr 
&&\sum_{ r , s , t } 
\sum_{ \alpha } B_{k \beta }  ~ D^R_{ij} (  ( rs ) \alpha ( rt )  )
  ~    C^{ \tau  , \Lambda, k  }_{ R , R , i , j } ~   
  (X_{ a_1}) {}^{ i_1}_{i_\alpha( 1 ) } \cdots X_{a_n}{}^{i_n}_{i_{\alpha(n)}} 
\eea 
The action of $ ( rs )$ on the upper indices and $ (rt)$ on the lower indices 
of the $ X$'s have  been  absorbed into a re-definition of $ \alpha $. 
Expanding 
\bea 
&& \sum_{ r , s , t } 
\sum_{ \alpha } B_{k \beta }  ~ D^R_{ik } (  ( rs ) )   D^R_{kl} (  \alpha ) 
D^{R}_{lj} (  ( rt )  )
  ~    C^{ \tau  , \Lambda, k  }_{ R , R , i , j } ~   
  (X_{ a_1}) {}^{ i_1}_{i_{ \alpha( 1 ) }  } \cdots 
( X_{a_n}) {}^{i_n}_{i_{\alpha(n)}} \cr 
&& =  ~ C_{ T }  ( \tau ) ~  \sum_{ \alpha  } B_{ k \beta }  C_{R R k l }^{ \Lambda, m_{ \Lambda} , \tau  } ~   
 D^R_{kl} (  \alpha ) ~ (X_{ a_1}) {}^{ i_1}_{i_{ \alpha( 1 ) }  }
 \cdots ( X_{a_n}) {}^{i_n}_{i_{\alpha(n)}} \cr 
&& = C_{ T } ( \tau ) ~ \cO^{ \Lambda, R , \beta , \mu , \tau } 
\eea 
In the second line we used (\ref{tauident}). 
We have thus proved the third line of (\ref{toshow})

%%%%%%%%%%%%%%%%%%%%%%%%%%%%%%%%%%%%%%%%%%

\section{Examples of generalised Casimirs for resolving multiplicity  }
\label{sec:examples} 
In this section, we explicitly 
calculate the action of 
the generalised Casimirs 
for some examples. 
We use the fact that 
the generalised Casimirs ($u(N)$ invariants) can be expressed in terms of 
the symmetric group or
the Brauer algebra, which is explained in 
the Appendix \ref{sec:UNtoFinitealgebra}. 

As we see in (\ref{restrictmultiplicity}) or 
(\ref{taugeneralised}) (or more generally (\ref{braueractiononQ})), 
the actions of the Casimirs can be effectively expressed 
by matrices whose sizes are given by the multiplicities. 
We explicitly obtain these matrices for some examples. 

Calculations in this section 
have heavily used the SAGE
\footnote{
http://www.sagemath.org/} interface 
to SYMMETRICA\footnote{
http://www.neu.uni-bayreuth.de/de/Uni\_Bayreuth
/Fakultaeten/1\_Mathematik\_Physik\_und\_Informatik\\
/Fachgruppe\_Informatik/prof\_diskrete\_algorithmen/en/research/SYMMETRICA/index.html} 
to get orthogonal matrix representations of 
the symmetric group and 
the MAXIMA \footnote{
http://maxima.sourceforge.net/
} for diagonalisation of matrices.

\subsection{One complex matrix : Brauer } 
In this subsection, we treat 
the Brauer basis of 
non-holomorphic operators constructed from 
one complex matrix, which is explained in section \ref{sec:ehsBrauer} 
and appendix \ref{sec:SymmetricBranchingOperator}. 
This basis is provided by the operator 
$Q^{\gamma}_{A,ij}$. 
We first construct a representation basis 
$|\gamma,A,m_{A},i\rangle$ 
representing the reduction of 
$\gamma$ to $A$, where $m_{A}$ labels states in 
$A$ and $i$ labels the multiplicity of this reduction.
We next express 
the generalised Casimir 
(\ref{c++-Brauer}) 
as a matrix whose size is 
given by the multiplicity. 

We consider the representation 
$\gamma=(k=2,\gamma^{+}=[1],\gamma^{-}=[1])$ of $B_{N}(3,3)$. 
This is the simplest case where the nontrivial multiplicity occurs. 
In this case the representation $A=([2,1],[2,1])$ 
of the subalgebra $ \mathbb C ( S_3 \times S_3 )$ 
appears with the multiplicity $2$ due to
$M_{A}^{\gamma}=2$ (see (3.13) in \cite{kr} 
 for the definition of $M_{A}^{\gamma}$). 
Because the dimension of the $A$ is $4$, there are $8$ 
states labelled by the $\gamma$ and the $A$. 
To identify these states, we use Cartan elements for 
$ \mathbb C ( S_3 \times S_3 )$, which 
are given by 
$X_{2}^{+}=(12)$, $X_{3}^{+}=(13)+(23)$, 
$X_{2}^{-}=(\bar{1}\bar{2})$ and 
$X_{3}^{-}=(\bar{1}\bar{3})+(\bar{2}\bar{3})$. 
As is reviewed in the Appendix \ref{sec:vershikokounkov}, 
%the Caltan elements should have eigenvalues 
%$(X_{2}^{+},X_{3}^{+})=(\pm1,\mp1)$ and 
%$(X_{2}^{-},X_{3}^{-})=(\pm1,\mp1)$ for 
%states corresponding to $A=([2,1],[2,1])$. 
four sets of eigenvalues 
of Cartan elements 
$(X_{2}^{+},X_{3}^{+},X_{2}^{-},X_{3}^{-})=
(1,-1,1,-1)$, $(1,-1,-1,1)$, $(-1,1,1,-1)$, 
$(-1,1,-1,1)$ correspond to $m_{A}=1,2,3,4$. 

Representation matrices for 
$\gamma=(k=2,\gamma^{+}=[1],\gamma^{-}=[1])$ are constructed 
in the Appendix \ref{sec:Brauerrep}, and we use them. 
For example, 
two orthonormalised states with eigenvalues 
$(D^{\gamma}(X_{2}^{+}),
D^{\gamma}(X_{3}^{+}),D^{\gamma}(X_{2}^{-}),
D^{\gamma}(X_{3}^{-}))=(1,-1,1,-1)$ can be found as 
\begin{eqnarray}
&& 
v_{1}=
\frac{1}{2\sqrt{5}}\left(
e_{1}+e_{3}
-e_{5}
-e_{6}+e_{7}+e_{9}-e_{11}-e_{12}-e_{13}-e_{14}-e_{15}-e_{16}
+e_{17}+e_{18}
\right) \cr
&&
v_{2}=
\frac{1}{6\sqrt{5}}\left(
4e_{1}-5e_{2}
+
4e_{3}-5e_{4}
-e_{5}
-e_{6}+4e_{7}-5e_{8}+4e_{9}
-5e_{10} \right. \cr 
&& \left. \qquad
+e_{11}+e_{12}+e_{13}+e_{14}+e_{15}+e_{16}
-2e_{17}-2e_{18}
\right) 
\label{1-11-1states}
\end{eqnarray}
where $e_{i}$ $(i=1,\cdots,18)$ are 
basis vectors of the representation $\gamma$ of $B_{N}(3,3)$. 
These two states can be written as $|\gamma,A,(1,-1,1,-1),i\rangle $ 
$(i=1,2)$. 

These two states should have 
different multiplicity labels, and 
we will see that they are resolved by eigenvalues of the generalised 
Casimir operator introduced in (\ref{c++-Brauer}). 

\begin{eqnarray}
T^{++-}\equiv \sum_{r\neq s=1}^{3}\sum_{t=4}^{6}(rs)
C_{r\bar{t}}
\label{generalised33}
\end{eqnarray}
Action of this operator on the two states 
are given by 
\begin{eqnarray}
\frac{1}{5}
\left(
\begin{array}{ccc}
4N+12 & -3N+6 \\
-3N+6 & -4N+3 \\
\end{array}
\right)
\end{eqnarray}
where $(ij)$ component represents 
$\langle v_{i}| T| v_{j} \rangle$. 
The eigenvalues are $-(3\pm \sqrt{4N^{2}+9})/2$. 
In this way, the action of the Casimir on the representation 
basis can be expressed by the $2\times 2$ matrix.

We also obtained another set of eigenvectors with 
eigenvalues $(1,-1,-1,1)$ corresponding to 
another state of $A=([2,1],[2,1])$. 
We found these vectors are also labelled 
by the same 
eigenvalues of $T^{++-}$. 
This fact is what is expected from the Schur's lemma. 
We can check that 
the Casimir (\ref{generalised33})
commutes with all elements in the subalgebra 
$ \mathbb C ( S_3 \times S_3 )$. 
Therefore the Casimir acts diagonally in the state labels $ m_{ \Lambda} $ 
as in (\ref{taugeneralised}).

\subsection{One Complex Matrix : Restricted Schur } 
In this subsection, 
another basis of one complex model, 
restricted schur basis, is investigated. 
This basis is given by $Q^{R}_{R_{1}R_{2};ij}$. 

We consider 
the case $ S_6 $ reducing to $ S_3 \times S_3 $ as a example.
In the case, 
we have 
the multiplicity is $2$ for 
$ R_1 = [2,1] , R_2 = [2,1] $ and $ R = [ 3,2,1]$ because 
the multiplicity is given by the Littlewood-Richardson 
coefficient, 
$g([2,1],[2,1];[3,2,1])=2$. 
There are $8$ states labelled by the $R_{1}$ and $R_{2}$ 
inside the representation $R$. 
To look for the states, we use 
Cartan elements for $\mathbb C (S_{3}\times S_{3})$, 
which are denoted by 
$(X_{2}^{L},X_{3}^{L},X_{2}^{R},X_{3}^{R})$. 
These $8$ states have eigenvalues
$(X_{2}^{L},X_{3}^{L})=(\pm1,\mp1)$ and
$(X_{2}^{R},X_{3}^{R})=(\pm1,\mp1)$. 
Representation matrices for $R=[3,2,1]$ 
can be obtained using the SYMMETRICA\footnote{
We use the function symmetrica.odg in SYMMETRICA 
which gives the symmetric group matrix elements in 
an orthonormal bases. 
}. 
The dimension of this representation is $16$. 
Let $e_{i}$ $(i=1,\cdots,16)$ be basis vectors of 
the representation matrices. 

Two orthonormalised eigenvectors with 
($D^{R}(X_{2}^{L}))$,$D^{R}(X_{3}^{L})$,$D^{R}(X_{2}^{R})$,$D^{R}(X_{3}^{R}))=(1,-1,1,-1)$
are found to be
\begin{eqnarray}
&&
\frac{\sqrt{2}}{4}
\left(e_{3}+\frac{1}{\sqrt{3}}e_{5}-\sqrt{5}e_{13}
-\sqrt{\frac{5}{3}}e_{15} \right) \cr
&&
\frac{\sqrt{10}}{8}
\left(
e_{3}+\frac{1}{\sqrt{3}}e_{5}
-\sqrt{3}e_{8}-\frac{3}{\sqrt{5}}e_{10}
-\frac{1}{\sqrt{5}}e_{13}
+\frac{1}{\sqrt{15}}e_{15}
\right)
\end{eqnarray}
These should be written as 
$|R,R_{1},R_{2},(1,-1,1,-1),i\rangle$, where $i$ takes $1,2$. 
The action of 
the Casimir $\sum_{s\neq r=1}^{3}\sum_{t=4}^{6}(rs)(rt)$ on these states 
are expressed by
the following $2\times 2$ matrix
\begin{eqnarray}
\left(
\begin{array}{ccc}
-4 & \sqrt{5} \\
\sqrt{5}& 1 \\
\end{array}
\right)
\end{eqnarray}
with eigenvalues $(3\pm 3\sqrt{5})/2$.
 
We also obtained another set of eigenvectors with 
eigenvalues $(1,-1,-1,1)$, and checked these are 
labelled by the same eigenvalues of the Casimir. 
Because the Casimir commutes with all elements of 
$S_{3}\times S_{3}$, it is consistent with 
the Schur's lemma.

\subsection{Two complex matrices : $U(2)$ covariant } 
In this subsection, we 
study the $\tau$ multiplicity which arises 
in holomorphic operators constructed from two complex matrices, 
which is explained in section \ref{sec:umsnII}. 
This multiplicity is related to the reduction 
$S_{n}\times S_{n}\rightarrow S_{n}$. 

We consider the 
case of $n=5$, which is the simplest case to see 
the multiplicity. 
The (inner) tensor  product 
between $[3,1,1]$ and itself can 
be decomposed into 
\begin{eqnarray}
[3,1,1]\otimes [3,1,1]
=[5]\oplus [4,1]\oplus 2[3,2]\oplus [3,1,1]
\oplus 2[2,2,1]\oplus [2,1,1,1]\oplus [1,1,1,1,1]
\end{eqnarray}
Here two irreducible representations 
$[3,2]$ and $[2,2,1]$ 
appear with the multiplicity $2$. 
From representation matrices for the representation of 
$[3,1,1]$ whose dimension is $6$, 
we construct 
$36\times 36$ matrices by 
taking the (inner) tensor product of them.  
To identify states corresponding to $[3,2]$ and $[2,2,1]$, 
we use Cartan elements for $\C(S_{5})$, which are denoted by 
$X_{i}$ ($i=2,\cdots,5$). 
Five components in the representation 
$[3,2]$ can be labelled by eigenvalues of the Cartan elements as
$(R^{R\otimes R}(X_{2}),R^{R\otimes R}(X_{3}),R^{R\otimes R}(X_{4}),R^{R\otimes R}(X_{5}))=(-1,1,0,2)$, $(1,-1,0,2)$, 
$(-1,1,2,0)$, $(1,-1,2,0)$, $(1,2,-1,0)$. 
These correspond to the five standard tableaux listed in table 
\ref{32S5standard}. 
On the other hand, 
five states 
in the representation $[2,2,1]$ 
are by 
$(-1,-2,1,0)$, $(-1,1,-2,0)$, $(1,-1,-2,0)$, $(-1,1,0,-2)$, 
$(1,-1,0,-2)$. 
See figure \ref{221S5standard} for standard tableaux 
corresponding to these sets of eigenvalues. 

Eigenvectors with $(1,2,-1,0)$, which corresponds 
to a state of $[3,2]$, can be found as 
\begin{eqnarray*}
&&u_1=\frac{1}{2\sqrt{3}}\left(
2e_{1}-e_{8}-e_{15}-e_{22}-e_{29}+2e_{36}
\right) \cr
&&u_2=\frac{1}{6\sqrt{2}}\left(
-2e_{1}+e_{8}+\sqrt{15}e_{10}
+e_{15}+\sqrt{15}e_{17}+\sqrt{15}e_{20}-e_{22}
+\sqrt{15}e_{27}-e_{29}
+2e_{36}
\right) 
\end{eqnarray*}
The action of the Casimir 
$\sum_{r\neq s\neq t}(rs)\otimes (rt)$ 
on these states can be expressed,  using (\ref{CHact}), 
 by the following matrix 
\begin{eqnarray}
\left(
\begin{array}{ccc}
6 & 0 \\
0& -4 \\
\end{array}
\right)
\end{eqnarray}
We also computed eigenvectors with $(1,-1,2,0)$ corresponding 
to another state in the representation $[3,2]$, 
which gave the same eigenvalues as the above. 
This is consistent with the Schur's lemma because 
the Casimir is invariant under the diagonal $S_{5}$. 

As for the representation $[2,2,1]$, 
we obtained orthonormalisd two eigenvectors with eigenvalues 
$(1,-1,0,-2)$ as 
\begin{eqnarray}
v_1&=&
\frac{1}{2\sqrt{33}}
\left(
\sqrt{10}e_{2}
+\sqrt{10}e_{7}
-\sqrt{5}e_{8}
+\sqrt{5}e_{15}
-2\sqrt{6}e_{19}
-2\sqrt{3}e_{20}\right.\cr
&&\left.
+\sqrt{5}e_{22},
+2\sqrt{3}e_{27}
-\sqrt{5}e_{29}
-\sqrt{10}e_{30}
-2\sqrt{6}e_{33}
-\sqrt{10}e_{35}\right) \cr
v_2&=&
\frac{\sqrt{11}}{44\sqrt{6}}
\left(
-\sqrt{30}e_{2}
+11\sqrt{2}e_{4}
-\sqrt{30}e_{7}
+\sqrt{15}e_{8}
+11e_{10}
-\sqrt{15}e_{15}
\right.
\cr
&&\left.
-11e_{17}
+11\sqrt{2}e_{18}
-5\sqrt{2}e_{19}
-5e_{20}
-\sqrt{15}e_{22}
+5e_{27} \right.
\cr
&&\left.
+\sqrt{15}e_{29}
+\sqrt{30}e_{30}
-5\sqrt{2}e_{33}
+\sqrt{30}e_{35}\right)
\end{eqnarray}
In this case, 
the action of the Casimir is given by 
\begin{eqnarray}
\frac{2}{11}\left(
\begin{array}{ccc}
-18 & 10\sqrt{6} \\
10\sqrt{6}& 7 \\
\end{array}
\right)
\end{eqnarray}
whose eigenvalues are $-6$ and $4$. 
We also obtained the same eigenvalues for 
eigenvectors with eigenvalues 
$(1,-1,-2,0)$ which is another states of 
$[2,2,1]$.

%%%%%%%%%%%%%%%%%%%%%%%%%%%%%%%%%%%%%%%%%%%%%%%%%%%%%%%%%%%%%%%%%%%%%%%%%%
%%%%%%%%%%%%%%%%%%%%%%%%%%%%%%%%%%%%%%%%%%%%%%%%%%%%%%%%%%%%%%%%%%%%%%%%%%

\section{Discussion  }

The enhanced symmetries are broken at non-zero coupling.  
The breaking should lead to Ward identities which 
control the mixing between the free field diagonal basis 
labels caused by the action of the one and higher loop 
Hamiltonians. The  mixing has been studied 
explicitly in the 1-loop case \cite{tom1loop}, where 
it was shown that the mixing of the $R$-label 
was limited by a simple rule of adding and removing a box 
of the Young diagram.  

Often the same sector of the theory admits different 
diagonal bases. The case of one complex matrix has 
been discussed at length. Another example in $ \cN=4 $ SYM 
is the holomprhic sector of $3$ complex matrices. 
We can diagonalise with the $ U(3) \times S_n $ covariant 
method of \cite{bhr2}, where the $S_n$ is dual to $U(N^2)$. 
Alternatively we have the method of \cite{bcd} 
which keeps $S_{n_1} \times S_{n_2} \times S_{n_3} $ 
manifest. The latter is Schur-Weyl dual to $ U(N) \times U(N) \times U(N)$. 
The transformation between the two bases should have a 
group theoretic meaning in terms of the relations between 
the different unitary groups involved.

\subsection{An extremality property } 
In the context of the holomorphic sector of a single matrix $X$ 
and the related half-BPS  LLM geometries, 
we can view the higher Casimirs as space-time charges
encoded in the asymptotics of the gravitational fields
 \cite{integinfo}. It is reasonable to  assume  that 
the existence of these charges  holds beyond the SUSY case
even though the available description uses the multipoles of  
 the $u$-function characterising the LLM geometries. 
In the case of one complex matrix, it was argued \cite{kr} that 
the Brauer basis has an  interpretation in terms of 
branes and anti-branes. Recall that the label $ \gamma $ is equivalent  to  
$ ( \gamma_+ , \gamma_- , k ) $ where $\gamma_+ $ is a partition of
$m-k$ and $ \gamma_- $ is a partition of $ n-k $, expressed as 
$ \g_+ \vdash m-k , \g_- \vdash n-k$. 
In particular it was argued  that $k =0 $ operators correspond to 
the ground state of a brane corresponding to $ R \vdash m , S \vdash n $. 
The higher $k$ operators are excited states. 
In support of this interpretation we now see that the $k=0$ operators 
obey a type of extremality condition.   For fixed $C_2 ( \gamma ) $,
as we increase $k$, the energy increases. So the $k=0$ states are minimal 
energy states for fixed higher charges.

\subsection{Reduced quantum mechanics  }

Much of our discussion can be carried over to the 
case of the reduced quantum mechanics coming from 
the complex matrices. As an example consider the case of 
a single complex matrix. In the reduced quantum mechanics,
 we can use our  $ X , X^{ \dagger} $ operators by
 replacing $ X \rightarrow A^{ \dagger} $ and 
 $ Y \rightarrow B^{\dagger} $. The main point that different diagonal bases 
 are related to different enhanced symmetries can be expressed in 
 the reduced models.

\subsection{Finite $N$ } 

The discussion of the Brauer algebra construction 
of gauge invariant operators has some interesting subtleties 
at finite $N$. Two things happen at finite $N$. The Brauer 
algebra $B_{N} ( m ,n ) $ 
 maps onto the commutant of $V^{\otimes m } \otimes \bar V^{ \otimes n } $
  but the map is not one-to-one. 
 This is a feature that is seen already in the case of 
 the duality between symmetric groups and the  unitary group action on 
 $V^{\otimes m } $. 
Unlike the symmetric group case,  
the Brauer algebra has structure constants that 
themselves depend on $N$. It ceases to be semi-simple 
 in the case of $ m+n > N $. This means there is no longer a non-degenerate 
trace on the algebra, which implies that the general construction of 
matrix units in the Appendix does not work.  
 A related consequence is that some of 
 the projectors constructed explicitly in \cite{kr} 
 become singular when $ m +n  > N $. The systematic treatment of these 
singularities for the general $ Q^{ \g }_{ A , i , j } $ 
is a problem  we leave for the future. In some special cases 
it is clear what the effects of finite $N$ are. Indeed for 
$k=0$, where the $ Q^{ \g }_{ A , i  j } $ 
 are ordinary projectors $P_{ R \bar S } $ 
 the cutoff is simply $ c_1( R ) + c_1( S ) \le N $, which 
was interpreted as an exclusion principle for brane-anti-brane
ground states \cite{kr}.  
 The finite $N$ effects in the restricted Schur description of 
 the gauge invariant operators in the $ X , X^{ \dagger} $ 
 sector are easier to handle \cite{bcd} since we do not have to 
 deal with the non-semisimplicity issue. Given that the same 
 space of operators is being described by $B_{ N} ( m , n ) $ and 
$ S_{ m + n } $, we can expect to find some information 
 about finite $N$ effects on the Brauer algebra using this set-up. 
For some mathematical work on finite $N $ effects in  
 algebra $B_N ( m ,n ) $ see \cite{martin}.
 An interesting future direction is to give a 
 complete finite $N$ discussion of the $ Q^{ \g}_{ A , i , j } $ 
 and to develop a  branes-anti-brane interpretation  
 of  the finite $N$ cutoffs.

\section{Summary and Outlook } 

The zero coupling limit of $ \cN =4 $ SYM with $U(N)$ symmetry  
has enhanced global symmetries. There is a $U(N)^{\times 4 } $
which acts on a  complex adjoint scalar field, leaving other fields 
unchanged, which can be constructed by the standard Noether procedure. 
 This is a symmetry of the classical dilatation operator, 
which is the Hamiltonian of radial quantisation.  The diagonal
$U(N)$ subgroup of this product is the global gauge symmetry. 
This leaves physical states invariant. This invariance condition
forces the local physical operators to be traces.  The full $U(N)^{\times 4} $ 
acts non-trivially on the space of gauge invariant operators. 
By the operator-state correspondence, these gauge invariant operators 
correspond to physical states. The $ U(N)^{\times 4} $ is a subgroup 
of $ (  U(N^2) )^{ \times 2 } $ which is also present in the theory.
These symmetries exist also for the full set of fundamental 
fields. There are are $8$ copies of $U(N)^{\times 4} \subset 
(  U(N^2) )^{ \times 2 } 
$, of which $4$
on bosonic fields and $4$ on fermionic fields.  
We have used generalised Casimirs constructed from the generators of 
the enhanced  symmetries in order to distinguish  the 
labels on bases for the complete set of operators at finite $N$
which diagonalise the two-point functions. 
Some of these Casimirs are rather standard, others are less 
familiar but nevertheless have the expected invariance properties. 
Schur-Weyl duality allows the expression of the Casimirs 
constructed from the Noetherian symmetries in terms of
dual algebras including the symmetric group algebra, Brauer 
algebras and generalisations.

A frequently asked question by students of AdS/CFT 
is how to see from the string theory of AdS the large 
$U(N)$  symmetry of the gauge theory. The usual reply from 
the teachers is that the $U(N)$ is not a dynamical symmetry 
acting on the physical spectrum, so we should not
 expect to see it in the string theory.
This paper revives the question in a refined form.
 How do we see the charges constructed from the enhanced symmetry 
 $ U(N)^{\times 4} $ from the dual string theory in the limit 
 of $g_s , \lambda  \rightarrow 0$ with ${ \lambda \over g_s } = N $ fixed ? 
Is there a two-dimensional worldsheet formulation
in this double scaling limit  which captures 
the finite $N$ effects such as the  cutoffs discussed as 
the stringy exclusion principle \cite{malstrom,jevram,horam}
or  the enhanced symmetries discussed here, and treats $g_s N $ 
as a deformation parameter that breaks the symmetries ? 
This would not be a conventional perturbative string 
but perhaps something more like Matrix string theory \cite{DVV} 
which contains $N$ as an integer parameter.

The idea of  studying the enhanced symmetry points of 
string theory as an approach  that makes optimal 
use of its hidden symmetries  is one that has come up several times 
in the past,e.g \cite{finiteal}. The hope is 
 that the enhanced symmetries 
can be used to organise the physics at the special points 
and also yield information away from the special points. 
Thanks to the  AdS/CFT duality, the 
zero coupling limit of Yang Mills gives a tractable 
situation where the power and usefulness of this strategy 
can be developed, explored and scrutinised for  lessons 
applicable to general string backgrounds. The explicit construction 
of the worldsheet perturbation expansion has been initiated in 
\cite{gopak}. Interesting work on the string theory intepretation 
of free Yang Mills has also been done in \cite{hagsun}. 

The simplest of the  Casimirs we have considered, 
which are   relevant to the diagonalisation of 
holomorphic operators have been used in discussions of 
how to extract detailed information about 
half-BPS solutions including extrenal black holes  \cite{integinfo} 
from the asymptotic fields in the dual spacetime. As explained in 
section \ref{sec:generalcase} we can apply the ideas described here to 
construct Casimirs which distinguish the complete set of states 
of a diagonal basis of free $\cN =4 $ SYM. 
This  gives an argument in favour  of 
the idea of integrability of $\cN =4 $ SYM,  in a set up that 
naturally incorporates multi-traces and finite $N$ effects.
As such it  supports the view that these charges contain 
complete information about black holes in $ AdS_5 \times S^5 $ 
spacetime beyond the half-BPS sector. 
Finding spacetime duals of these enhanced symmetry charges
 is an important problem. Possibly this would require 
 taking some limit of large quantum 
 numbers where supergravity  or semiclassical brane physics 
can be compared  with weak coupling gauge theory. Alternatively 
a rough qualitative comparison of the number of charges 
constructed here with those constructible from gravity in spacetime 
would be interesting.

\vskip.7in 

{ \bf Acknowledgements } We thank Tom Brown, Rajesh Gopakumar, 
 Horatiu Nastase, 
Gabriele Travaglini, David Turton  for discussions. 
Special thanks to Tom Brown for introducing us to the mathematical software 
SAGE, which we used extensively especially as 
an interface to SYMMETRICA.   YK is supported 
 by an STFC grant PP/D507323/1  ``String theory, gauge theory and gravity''. 
 SR is supported by an STFC Advanced Fellowship and STFC grant PP/D507323/1. 

\vskip1in

\appendix

%%%%%%%%%%%
%%%%%%%%%%%

\section{Casimirs : From $ U(N)$ expressions to
 expressions in terms of finite algebras   } 
\label{sec:UNtoFinitealgebra}
In this section, we express Casimirs of the $U(N)$ group
in terms of the symmetric group or the Brauer algebra. 
This relation essentially comes from the Schur-Weyl  duality. 

\subsection{$V^{\otimes n}$}
Let $E^{i}_j $ be the generator of $u(N)$. 
The action of it on the space of 
the fundamental representation $V$ is given by 
\bea 
E^{i}_j v^{k  } =  \delta^{k }_{j} v^{i} 
\label{UNaction}
\eea
where $v^{i}$ denotes the standard basis for the fundamental 
representation $V$. 
Consider the tensor space $V^{\otimes n}$. 
The action of the generator
can be extended to this tensor space 
as $\sum_{s=1}^{n}\rho_{s}(E)$, where $\rho_{s}(E)$ represents 
the action on the $s$-th space. 
We also use a notation $\rho_{\underline{n}}(E) \equiv 
 \sum_{s=1}^{n}\rho_{s}(E) $. 
The quadratic Casimir on this space can be computed as
\begin{eqnarray}
\hat{C}_{2}
&=&\sum_{r,s=1}^{n}\rho_{r}(E^{i}_{j})\rho_{s}(E^{j}_{i})\cr
&=&\sum_{r=1}^{n}\rho_{r}(E^{i}_{j}E^{j}_{i})
+\sum_{r\neq s} \rho_{r}(E^{i}_{j}) \rho_{s}(E^{j}_{i})
\cr
&=&Nn+
\sum_{r\neq s}(rs)
\label{quadraticVn}
\end{eqnarray}
Here we have assumed the action on 
$v^{i_{1}}\otimes \cdots \otimes v^{i_{n}}$, and 
the following equations were used. 
\begin{eqnarray}
&&(E^{i}_{j}E^{j}_{i}) v^{k}=Nv^{k}\cr
&&
(E^{i}_{j}\otimes E^{j}_{i}) v^{k_{r}}\otimes v^{k_{s}}=
v^{k_{s}}\otimes v^{k_{r}}=(rs)v^{k_{r}}\otimes v^{k_{s}}
\end{eqnarray}
On an irreducible space 
$p_{R}V^{\otimes n}$, we recover a known relation 
\bea 
C_2 ( R ) = Nn +2 { \chi_R ( T_2 ) \over d_R }  
\eea
$T_{2}$ is a sum of the transpositions.

%%%%%%%%%%%%%%%%%%%

\subsection{$V^{\otimes m}\otimes \bar{V}^{\otimes n}$ }
\label{CasimirBrauer}

We describe some relevant properties of the Casimirs used in 
section \ref{sec:ehsBrauer}. 
The action of $u(N)$ on the dual space is 
\bea 
E^{i}_j \bar v^{r  } = - \delta^{i r } \bar v^{j} 
\eea
Using this action we verify that $ v^p \bar v^p $ is invariant under $U(N)$. 
On the space $V^{\otimes m}\otimes \bar{V}^{\otimes n}$ 
the action of $U(N)$ is dual to the action of 
the Brauer algebra (Schur-Weyl duality). 
So we can express $U(N)$ Casimirs in terms of 
elements of the Brauer algebra. 
Calculations are almost parallel to the case of 
$V^{\otimes n}$. 
In this case we will need 
\begin{eqnarray}
&&(E^{i}_{j}E^{j}_{i}) \bar{v}^{k}=N\bar{v}^{k}
\cr
&&
(E^{i}_{j}\otimes E^{j}_{i})
 \bar{v}^{k_{r}}\otimes \bar{v}^{k_{s}}=
\bar{v}^{k_{s}}\otimes \bar{v}^{k_{r}}
=(\bar{r}\bar{s}) \bar{v}^{k_{r}}\otimes \bar{v}^{k_{s}}
\cr
&&
(E^{i}_{j}\otimes E^{j}_{i}) v^{k_{r}}\otimes \bar{v}^{k_{s}}=
-\delta^{k_{r}k_{s}}\sum_{p}v^{p}\otimes \bar{v}^{p}
=-C_{r\bar{s}}v^{k_{r}}\otimes \bar{v}^{k_{s}}
\end{eqnarray}
$C$ is the contraction which is a linear map 
from $V\otimes \bar{V}$ to $V\otimes \bar{V}$. 
The quadratic Casimir on the space 
$V^{\otimes m}\otimes \bar{V}^{\otimes n}$
can be computed by 
\begin{eqnarray}
\hat{C}_{2}
&=&\sum_{r,s=1}^{m+n}\rho_{r}(E^{i}_{j})\rho_{s}(E^{j}_{i})\cr
&=&\sum_{r=1}^{m+n}\rho_{r}(E^{i}_{j}E^{j}_{i})
+\sum_{r\neq s} \rho_{r}(E^{i}_{j}) \rho_{s}(E^{j}_{i})
\cr
&=&(m+n)N+
\sum_{r\neq s=1}^{m}(rs)+\sum_{r\neq s=m+1}^{m+n}(\bar{r}\bar{s})
-2\sum_{r=1}^{m}\sum_{s=m+1}^{m+n}C_{r\bar{s}}
\label{c_{2}onVbarV}
\end{eqnarray}
This is a central element of the Brauer algebra. 

We next consider 
another kind of Casimir invariant which plays an important 
role in this paper. 
We introduce $U(N)\times U(N)$, 
where the first $U(N)$ acts on $V$ 
and the second one acts on $\bar{V}$. 
This $U(N)\times U(N)$ action on 
$V^{\otimes m}\otimes \bar{V}^{\otimes n}$ is 
dual to the action of 
$S_{m}\times S_{n}$. 
For example, we are interested in 
$\rho_{\underline{m}}(E^{i}_{j})
\rho_{\underline{m}}(E^{j}_{k})
\rho_{\underline{n}}(E^{k}_{i})$, which can be expressed as  
\begin{eqnarray}
\hat{C}_{++-}&=&
\sum_{r,s=1}^{m}\sum_{t=m+1}^{m+n}
\rho_{r}(E^{i}_{j})
\rho_{s}(E^{j}_{k})
\rho_{t}(E^{k}_{i}) \cr
&=&\sum_{r=1}^{m}\sum_{t=m+1}^{m+n}
\rho_{r}(E^{i}_{j}E^{j}_{k})
\rho_{t}(E^{k}_{i})
+\sum_{r\neq s=1}^{m}\sum_{t=m+1}^{m+n}
\rho_{r}(E^{i}_{j})
\rho_{s}(E^{j}_{k})
\rho_{t}(E^{k}_{i})  \cr
&=&-N\sum_{r=1}^{m}\sum_{t=m+1}^{m+n}C_{r\bar{t}}
-\sum_{r\neq s=1}^{m}\sum_{t=m+1}^{m+n}(rs)C_{r\bar{t}} \cr
&=&-N\sum_{r=1}^{m}\sum_{t=m+1}^{m+n}\left(
1+\frac{1}{N}\sum_{s(\neq r)=1}^{m}(rs)
\right)C_{r\bar{t}}
\end{eqnarray}
This is an element of  the Brauer algebra $ B_N(m,n)$ which  
is invariant under the subalgebra $\C(S_{m}\times S_{n})$.
Hence it can be used to distinguish between different 
multiplicity labels of representations $A$ 
of $\C(S_{m}\times S_{n})$ in an irrep $ \g $ of  $ B_N(m,n)$, 
and is  diagonal in the labels $m_A$ of states in $A$
\bea 
\la \g , A' , m'_{A' } , i | \hat C_{++-} | \g , A , m_{A } , j \ra 
= \delta_{ A' A } \delta_{ m_A m'_A } (  C_{++-} )_{ij}  
\eea 
This is demonstrated in section \ref{sec:examples}. 
A basis in the multiplicity space can be chosen to diagonalise 
$\hat C_{++-} $ and this is used in (\ref{pf223}).

\subsection{Casimirs for $V^{\otimes (m+n)}$}
\label{casimir-rest} 
We describe some relevant properties of the Casimirs 
used in section \ref{ehsrestschu}.
The quadratic Casimir $\rho(E^{i}_{j})\rho(E^{j}_{i})$ on the space 
$V^{\otimes (m+n)}$ is 
\begin{eqnarray}
\hat{C}_{2}
&=&\sum_{r,s=1}^{m+n}\rho_{r}(E^{i}_{j})\rho_{s}(E^{j}_{i})\cr
&=&\sum_{r=1}^{m+n}\rho_{r}(E^{i}_{j}E^{j}_{i})
+\sum_{r\neq s=1}^{m+n} \rho_{r}(E^{i}_{j}) \rho_{s}(E^{j}_{i})
\cr
&=&N(m+n)+
\sum_{r\neq s=1}^{m+n}(rs)
\end{eqnarray}

We next consider invariants constructed from 
$U(N)\times U(N)$, where the first $U(N)$ acts on 
$V^{\otimes m}$ and the second one acts on $V^{\otimes n}$.
The action of the $U(N)\times U(N)$ is dual to 
$S_{m}\times S_{n}$. 
\begin{eqnarray}
\hat{C}_{112}&=&
\sum_{r,s=m+1}^{m+n}\sum_{t=1}^{m}
\rho_{r}(E^{i}_{j})
\rho_{s}(E^{j}_{k})
\rho_{t}(E^{k}_{i}) \cr
&=&\sum_{r=m+1}^{m+n}\sum_{t=1}^{m}
\rho_{r}(E^{i}_{j}E^{j}_{k})
\rho_{t}(E^{k}_{i})
+\sum_{r\neq s=m+1}^{m+n}\sum_{t=1}^{m}
\rho_{r}(E^{i}_{j})
\rho_{s}(E^{j}_{k})
\rho_{t}(E^{k}_{i})  \cr
&=&N\sum_{r=m+1}^{m+n}\sum_{t=1}^{m}(rt)
+\sum_{r\neq s=m+1}^{m+n}\sum_{t=1}^{m}
(rs)(rt) \cr
&=&N\sum_{r=m+1}^{m+n}\sum_{t=1}^{m}\left(
1+\frac{1}{N}\sum_{s(\neq r)=m+1}^{m+n}(rs)
\right)(rt)
\end{eqnarray}
This 
is an element of $ \C  ( S_{m+n} )  $ which is  invariant under 
the subalgebra $\C(S_{m}\times S_{n})$. It is diagonal in the 
state labels of the subalgebra, and mixes the multiplicity labels
(see equation (\ref{restrictmultiplicity})). 
A diagonalising basis in the multiplicity space can be chosen. 
This is demonstrated in section  \ref{sec:examples} and is used in 
section \ref{PfssymmXX*}.

%%%%%%%%%%%%%%%%%%%%%%%%%%%%%%%%%%%%%%%

\subsection{$U(N^{2})$}
\label{sec:CasimirUN2}
Here we express Casimirs constructed  $U(N^{2})$ along with 
$U(N) \times U(N)$, 
in terms of the symmetric groups. 

Let $E^{jk}_{lm}$ be the generator of $U(N^{2})$ 
\begin{eqnarray}
E^{jk}_{lm}v^{p}_{q}=\delta^{p}_{m}\delta^{j}_{q}v^{k}_{l}
\end{eqnarray}
and $(E_{\cal{L}})^{l}_{j}$ and $(E_{\cal{R}})^{m}_{k}$ 
be the generators of the left and the right action of 
$U(N)$ 
\begin{eqnarray}
&&(E_{\cal{L}})^{l}_{j}v^{p}_{q}=\delta^{p}_{j}v^{l}_{q}\cr
&&(E_{\cal{R}})^{m}_{k}v^{p}_{q}=\delta^{m}_{q}v^{p}_{k}
\end{eqnarray}
The quadratic Casimirs of these actions of 
$U(N)$ are 
\begin{eqnarray}
&&
\hat{C}_{2}^{\cal{L}}
=\sum_{r,s=1}^{n}\rho_{r}((E_{\cal{L}})^{i}_{j})
\rho_{s}((E_{\cal{L}})^{j}_{i})
=Nn+
\sum_{r\neq s}(rs)_{\bold{u}} \cr
&&
\hat{C}_{2}^{\cal{R}}
=\sum_{r,s=1}^{n}\rho_{r}((E_{\cal{R}})^{i}_{j})
\rho_{s}((E_{\cal{R}})^{j}_{i})
=Nn+
\sum_{r\neq s}(rs)_{\bold{l}}
\end{eqnarray}
The calculations are completely similar to
(\ref{quadraticVn}). 
$(rs)_{\bold{u}}$ and $(rs)_{\bold{l}}$ 
are the actions on the upper and lower indices of 
$v^{p_{r}}_{q_{r}}\otimes v^{p_{s}}_{q_{s}}$, respectively.  

The quadratic Casimir for $U(N^{2})$ can be computed as 
\begin{eqnarray}
\sum_{r,s=1}^{n}\rho_{r}(E^{jk}_{lm}) 
\rho_{s}(E^{lm}_{jk}) 
&=&\sum_{r=1}^{n}\rho_{r}(E^{jk}_{lm}E^{lm}_{jk})
+\sum_{r\neq s}\rho_{r}(E^{jk}_{lm})
\rho_{s}(E^{lm}_{jk})  \cr
&=&\sum_{ r \ne s } (rs)_{\bl} (rs)_{\bu } + N^2 n 
\end{eqnarray}

We are interested in  invariants such as 
$\rho( E^{jk}_{lm})
\rho((E_{\cal{L}})^{m}_{k})
\rho((E_{\cal{R}})^{l}_{j}) $ which are  constructed from  $U(N^2)$ 
generators and which is invariant under the 
$ U(N) \times U(N)$ subalgebra.  
It is also 
expressed in terms of the symmetric groups as 
\begin{eqnarray}\label{dervnnn} 
&&
\sum_{r,s,t=1}^{n}\rho_{r}( E^{jk}_{lm})
\rho_{s}((E_{\cal{L}})^{m}_{k})
\rho_{t}((E_{\cal{R}})^{l}_{j}) \cr
&=&
\sum_{r\neq s\neq t}\rho_{r}( E^{jk}_{lm})
\rho_{s}((E_{\cal{L}})^{m}_{k})
\rho_{t}((E_{\cal{R}})^{l}_{j})
+
\sum_{r\neq t}
\rho_{r}( E^{jk}_{lm}(E_{\cal{L}})^{m}_{k})
\rho_{t}((E_{\cal{R}})^{l}_{j}) \cr
&&
+
\sum_{r\neq s}^{n}\rho_{r}( E^{jk}_{lm}(E_{\cal{R}})^{l}_{j})
\rho_{s}((E_{\cal{L}})^{m}_{k})
+
\sum_{r}\rho_{r}
( E^{jk}_{lm}(E_{\cal{L}})^{m}_{k}(E_{\cal{R}})^{l}_{j})\cr
&=&
 \sum_{ r \ne s \ne t } (rs)_{\bu}  (rt)_{\bl}  
 + N \sum_{ r\ne s } (rs)_{\bl} + N \sum_{ r\ne s }  (rs)_{\bu}  
 + \sum_{ r \ne s } (rs)_{\bl} (rs)_{\bu } + N^2 n 
\end{eqnarray}
For example, the first term was calculated as 
\begin{eqnarray}
&&
(E^{jk}_{lm}\otimes
(E_{\cal{L}})^{m}_{k}\otimes
(E_{\cal{R}})^{l}_{j})
v^{p_{r}}_{q_{r}}\otimes  v^{p_{s}}_{q_{s}}\otimes v^{p_{t}}_{q_{t}}\cr
&=&
v^{p_{s}}_{q_{t}} \otimes  v^{p_{r}}_{q_{s}} \otimes  v^{p_{t}}_{q_{r}}\cr
&=&
(rs)_{{ \bf u }  }(rt)_{{\bf l  }}  ~
v^{p_{r}}_{q_{r}}\otimes  v^{p_{s}}_{q_{s}} \otimes
v^{p_{t}}_{q_{t}}
\end{eqnarray}
This leads to the result (\ref{generalisedUN2}). 
The expression in the last line of (\ref{dervnnn}) is 
in $ \C ( S_n )  \times \C ( S_n ) $ and is invariant under the 
diagonal subalgebra $ \C ( S_n ) $. Such elements 
 can distinguish multiplicity 
labels for $S_n$ irreps. $\Lambda$ in the tensor products $R_1 \otimes R_2  $. 
This is illustrated in section \ref{sec:examples} 
and used in \ref{sec:umsnII}.

%%%%%%%%%%%%%%%%%%%%%%%%%%%%%%%%%%%%%%%%%%%%%%%%
\section{$Q$-operators  and the algebraic Brauer construction of 
matrix units }
\label{sec:SymmetricBranchingOperator}

In this section, 
we give an algebraic expression of the 
symmetric branching operator, which 
was proposed to give a complete set of gauge 
invariant operators constructed from 
$X$ and $X^{\dagger}$ \cite{kr}. 
An introduction of the Brauer algebra is given in the section 3 
in \cite{kr}. 

Representation matrix elements  of a Brauer algebra element $b$ 
are  denoted as 
\begin{eqnarray}
D_{IJ}^{\gamma}(b)=\langle \gamma,I |b|\gamma,J \rangle 
\quad 1 \le I,J \le d_{\gamma} 
\end{eqnarray}
where 
$I$ labels states in an irreducible representation $\gamma$ 
of the Brauer algebra, and 
$d_{\gamma}$ is the dimension of $\gamma$. 
The character is given by the trace of the matrix element 
\begin{eqnarray}
\sum_{I}D_{II}^{\gamma}(b)=
\sum_{I}\langle \gamma,I |b|\gamma,I \rangle =
\chi_{\gamma}(b) 
\end{eqnarray}
We define 
\begin{eqnarray}\label{matunits} 
P_{IJ}^{\gamma}\equiv t_{\gamma}
\sum_{i }D_{JI}^{\gamma}(b_i )b_i ^{\ast}, 
\label{PIJgamma}
\end{eqnarray}
which satisfies \cite{ramthesis}
\begin{eqnarray}
P_{IJ}^{\gamma}P_{KL}^{\gamma^{\prime}}
=\delta^{\gamma\gamma^{\prime}}\delta_{JK}P_{IL}^{\gamma}. 
\end{eqnarray}
$b^{\ast}$ is defined as 
a dual element of $b$ by $tr_{m,n}(bb^{\ast})=1$. 
For more details, 
see \cite{ramthesis} or the section 3 in \cite{kr}. 
The trace of $P_{IJ}^{\gamma}$ gives the central Brauer projector
\begin{eqnarray}
P^{\gamma}=\sum_{I}P_{II}^{\gamma}
=t_{\gamma}\sum_{i}\chi^{\gamma}(b_{i})b_{i}^{\ast}. 
\end{eqnarray}
The $ P_{IJ}^{\gamma}$ are elements of the algebra 
called {\it matrix units}. They can be identified 
with $ | \gamma , I \ra \la \gamma , J |$. The $ b_i^*$ is 
defined as the dual under a trace, i.e $ tr ( b_i b_j^* ) = \delta_{ij} $ 
\cite{ramthesis}.   
When the trace is in $ V^{ \otimes m } \otimes { \bar V }^{ \otimes n } $, 
the normalisation factor is $ t_{\gamma } = Dim \gamma$, which is 
the dimension of the $U(N)$ representation corresponding to the 
 Young diagram $ \gamma $, having positive row lengths 
adding to $m$ and negative row lengths adding to $n$. 

The matrix unit can be used to get a representation matrix 
from a character as 
\begin{eqnarray}
\chi^{\gamma}(P^{\gamma^{\prime}}_{IJ}b)
&=&
t^{\gamma^{\prime}}
\sum_{b^{\prime}}D_{JI}^{\gamma^{\prime}}(b^{\prime})
\chi^{\gamma}(b^{\prime \ast}b) \cr
&=&
t^{\gamma^{\prime}}
\sum_{b^{\prime}}D_{JI}^{\gamma^{\prime}}(b^{\prime})
D^{\gamma}_{KL}(b^{\prime \ast})D^{\gamma}_{LK}(b) \cr
&=&D_{JI}^{\gamma}(b)\delta^{\gamma\gamma^{\prime}}
\label{characterandrep}
\end{eqnarray}
where we have used the orthogonality of representation matrices 
for the Brauer algebra
\begin{eqnarray}
\sum_{b}D^{\gamma}_{IJ}(b)D^{\gamma^{\prime}}_{KL}(b^{\ast})
=\frac{1}{t^{\gamma}}\delta_{JK}\delta_{IL}
\end{eqnarray}

In our previous paper \cite{kr}, we have 
defined the 
$Q$-operator (symmetric branching operator) 
which has the following properties 
\bea 
&& Q^{\gamma}_{A,ij} Q^{\gamma^{\prime} }_{B,kl} = 
  \delta_{\gamma \gamma^{\prime}  } \delta_{AB}
 \delta_{jk} Q^{\gamma}_{A,il} 
 \label{defQ1}
 \\
&&  tr_{m,n} (  Q^{\gamma}_{A,ij}   )  = \delta_{ij} d_A  Dim \gamma  
\label{defQ2}
\\
&& h Q^{\gamma}_{A,ij}  h^{-1}= Q^{\gamma}_{A,ij}  
\qquad h \in \mathbb{C}  (S_m \times S_n)
\label{defQ3}
\eea 
$A$ labels an irreducible representation of $\mathbb{C}  (S_m \times S_n)$.
The indices $i,j$ are labels to distinguish 
different copies of the representation $A$ in the representation $\gamma$.

We now introduce the branching coefficient as 
\begin{eqnarray}
B_{\gamma , I ; A ,  m_A  , i }\equiv \langle \gamma, I 
|\gamma \rightarrow A,m_{A},i \rangle 
\label{branching}
\end{eqnarray}
The basis $|\gamma \rightarrow A,m_{A},i \rangle $ 
represents the decomposition of $\gamma$ in terms of 
the subalgebra $\mathbb{C}  (S_m \times S_n)$, and 
$m_A$ labels states in $A$ obeying $0\le m_{A}\le d_{A}$.  
We can always choose an orthogonal basis 
\begin{eqnarray}
\langle \gamma \rightarrow B ,m_{B}, k
|\gamma \rightarrow A ,m_{A}, j\rangle 
=\delta_{AB}\delta_{m_{A}m_{B}}\delta_{jk}
\end{eqnarray}
Using this orthogonality and the completeness of 
$|\gamma ,I \rangle $, we can show 
\begin{eqnarray}
\sum_{I}  (  B_{\gamma  , I; A , m_A  ,i_A })^{\dagger}
B_{\gamma , I ; B , m_B , i_B}
&=&
\sum_{I}
\langle \gamma \rightarrow A,m_{A},i_A  | \gamma, I \rangle 
\langle \gamma, I |\gamma \rightarrow  B ,m_{B},i_B \rangle \cr
&=& \delta_{AB } \delta_{m_A m_B } \delta_{i_A i_B }
\label{orthogonalitybranching}
\end{eqnarray}

The operator $Q_{A,ij}^{\gamma}$ can be expressed in terms of  
the operator (\ref{PIJgamma}) and 
the branching coefficient (\ref{branching}) as 
\begin{eqnarray}
Q_{A,ij}^{\gamma}
&=&\sum_{m_{A},I,J}
B_{\gamma I;A,m_{A},i}^{\dagger}P_{IJ}^{\gamma}B_{\gamma J;A,m_{A},j}
\label{QbyPB}
\end{eqnarray}
This expresses  $  Q_{A,ij}^{\gamma} $ in terms of Brauer  algebra 
elements.
It is easy to show (\ref{defQ1}) if we use 
the orthogonality of the branching coefficient 
(\ref{orthogonalitybranching}).
(\ref{defQ2}) can be shown as 
\begin{eqnarray}
tr_{m,n}(Q_{A,ij}^{\gamma})
&=&
\sum_{I,J}tr_{m,n}(P_{IJ}^{\gamma})
\sum_{m_{A}}
B_{\gamma I;A,m_{A},i}^{\dagger}
B_{\gamma J;A,m_{A},j}
\cr
&=& 
Dim\gamma 
\sum_{m_{A},I}
B_{\gamma I;A,m_{A},i}^{\dagger}
B_{\gamma I;A,m_{A},j}
\cr
&=&
Dim\gamma ~\delta_{ij} \sum_{m_A } \delta_{m_{A}m_{A}}
\cr
&=&
Dim\gamma ~ d_{A} ~ \delta_{ij}
\end{eqnarray}
To show the second line we have used 
$tr_{m,n}(P_{IJ}^{\gamma})=Dim\gamma\delta_{IJ}$ which 
follows from the Schur-Weyl duality for 
$ V^{ \otimes m } \otimes { \bar V}^{\otimes n } $.

Another useful property of $Q_{A,ij}^{\gamma}$ 
 is 
\bea 
(b) ~Q_{A,ij}^{\gamma} = \sum_k C^{\gamma}_{A,ki}(b) ~Q_{A,kj}^{\gamma}
 = { 1 \over d_A } \sum_{m_A , k } \la \g , A , m_A , k | (b) | \gamma , A , m_A , i \ra 
 ~   Q^{\gamma }_{A , k j }  
\label{braueractiononQ}
\eea 
Here $(b) $ is an element which commutes with any element 
in $ \mathbb C (S_{m}\times S_{n})$. 
This uses a related property of $ P^{\gamma}_{IJ}$ 
\bea 
b   P^{\gamma}_{IJ} = \sum_K D^{ \gamma }_{ KI }  ( b )  P^{\gamma}_{K J} 
\eea 

%%%%%%%%%%%%%%%%%%%%%%%

\subsection{Restricted characters for Brauer algebra }

Let us define the 
{\it restricted character} for the Brauer algebra
\begin{eqnarray}
\chi^{\gamma}_{A,ij}(b)
&\equiv&\sum_{m_{A},I,J}
B_{\gamma I;A,m_{A},i}^{\dagger}D^{\gamma}_{JI}(b)
B_{\gamma J;A,m_{A},j}\cr
&=&
\sum_{m_{A},I,J}
\langle \gamma \rightarrow A,m_{A},i |\gamma, I\rangle 
\langle \gamma,I |b|\gamma,J \rangle 
\langle \gamma, J |\gamma \rightarrow A,m_{A},j \rangle  
\cr
&=&
\sum_{m_{A}}
\langle \gamma \rightarrow A ,m_{A}, i|
b|\gamma \rightarrow A ,m_{A}, j\rangle 
\end{eqnarray}
The expression given by (\ref{QbyPB}) can be re-written  as 
\begin{eqnarray}
Q_{A,ij}^{\gamma}
&=&\sum_{m_{A}}
B_{\gamma I;A,m_{A},i}^{\dagger}P_{IJ}^{\gamma}B_{\gamma J;A,m_{A},j} \cr
&=&
t_{\gamma} \sum_{ b } 
\chi^{\gamma}_{A,ij}(b)
b^{\ast}
\label{restrictedBrauer}
\end{eqnarray}
The terminology ``restricted 
character''  has been used for analogous objects constructed 
in the reduction of $ S_{ m+ n } $ to $ S_m \times S_n$ 
in the context of strings attached to giant gravitons. We will 
elaborate on this below. 
Conversely the restricted characters can be expressed in terms of 
the $Q$ operators 
\begin{eqnarray}
\chi^{\gamma}_{A,ij}(b)
=\chi^{\gamma}(Q_{A,ji}^{\gamma}b)
\end{eqnarray}
This can be easily derived using 
(\ref{characterandrep}) and (\ref{QbyPB}). 
We can also show
\begin{eqnarray}
\chi^{\gamma}(b)=\sum_{A,i}\chi_{A,ii}^{\gamma}(b)
\end{eqnarray}

The following expression which is isomorphic to 
the algebraic expression is also useful to understand 
relations we have derived 
\begin{eqnarray}
Q^{\gamma}_{A,ij}=
\sum_{m}
|\gamma \rightarrow  A ,m,i \rangle 
\langle \gamma \rightarrow A,m, j  |
\end{eqnarray}
%and
%\begin{eqnarray}
%P^{\gamma}_{IL}=
%|\gamma ,I \rangle 
%\langle \gamma, L  |
%\end{eqnarray}

%%%%%%%%%%%%%%%%%%%%%%%%%%%%%%%%

\subsection{Symmetric elements in terms of Matrix units.  } 

Since 
$\sum_{h\in S_{m}\times S_{n}}h b h^{-1}$
commutes with any elements in the subalgebra 
$ \mathbb C (S_{m}\times S_{n})$, 
it can be expressed as a linear combination of $Q^{\gamma}_{A,ij}$.
\begin{eqnarray}\label{bsymexp} 
(b)\equiv \frac{1}{m!n!}
\sum_{h\in S_{m}\times S_{n}}h b h^{-1}
=C^{\gamma}_{A,ij}{}_{(b)}Q^{\gamma}_{A,ij}
\label{inverseQ}
\end{eqnarray}

Multiplying by $Q^{\gamma^{\prime}}_{ B, kl} $ on both sides of (\ref{bsymexp})
and using 
\bea 
&& Q^{\gamma}_{A,ij} Q^{\gamma^{\prime}}_{B,kl} =  
\delta_{\gamma \gamma^{\prime}} \delta_{AB}
 \delta_{jk} Q^{\gamma}_{A,il}\cr  
&&  tr (  Q^{\gamma}_{A,ij}   )  = \delta_{ij} d_A  Dim \gamma 
\eea 
we find 
the coefficient 
$C^{\gamma}_{A,ij}{}_{(b)}$ can be obtained as 
\begin{eqnarray}
C^{\gamma}_{A,ij}{}_{(b)}
&=&
\frac{1}{d_{A}Dim\gamma}tr_{m,n}
((b) Q^{\gamma}_{A,ji}) 
\cr
&=&
\frac{1}{d_{A}Dim\gamma}
\sum_{\gamma^{\prime}}
Dim\gamma^{\prime}
\chi_{\gamma^{\prime}}
((b) Q^{\gamma}_{A,ji}) 
\cr
&=&
\frac{1}{d_{A}}
\chi_{\gamma}
((b) Q^{\gamma}_{A,ji}) 
\cr
&=&
\frac{1}{d_{A}}
\chi^{\gamma}_{A,ij}
((b)) 
\cr
&=&
\frac{1}{d_{A}}
\chi^{\gamma}_{A,ij}(b) 
\end{eqnarray}
We have used the Schur-Weyl duality to get the second line, and 
the following property for the restricted character 
for the Brauer algebra defined in 
(\ref{restrictedBrauer})
has been used to derive the last equality 
\begin{eqnarray}
\chi^{\gamma}_{A,ij}
(h b h^{-1}) =\chi^{\gamma}_{A,ij}(b) ,\quad 
h\in \C(S_{m}\times S_{n} )
\end{eqnarray}

\subsection{Restricted Schur Operators for $ X , X^{ \dagger} $} 

The restricted Schur operators of \cite{bbfh,robgtsti,robgtstii,robgtstiii} 
can be described using the language developed above. This highlights 
the similarities  between the Brauer construction and the symmetric group 
construction,  explained  in broad outline 
at the beginning of section \ref{PfssymmXX*}. 

Analogous to (\ref{matunits}) the matrix units for $S_{m+n} $ are 
\bea 
P^R_{IJ} = {d_{R} \over (m+n)! } \sum_{ \sigma \in S_{m+n}  }
 D^{R}_{ JI}  ( \sigma  )  ~ \sigma^{-1} 
=   {d_{R} \over (m+n)! } \sum_{ \sigma \in S_{m+n}  }
 D^{R}_{ IJ }  ( \sigma  )  ~ \sigma 
\eea 
where we used 
$  D^{R}_{ JI  }  ( \sigma^{-1}   ) =  D^{R}_{ IJ   }  ( \sigma   )$. 
The analog of the $Q^{ \gamma}_{ \alpha , \beta ; i ,j } $ 
is $ Q^{ R }_{ R_1 , R_2 ; i , j } $ which can be written as 
\bea 
 Q^{ R }_{ R_1 , R_2 ; i , j }
 =    B_{  R_1 , R_2 , m_{R_1} , m_{ R_2 } ,  i ; R , I  }  P^R_{ IJ } B_{ R , J ; R_1 , R_2 , m_{R_1} , m_{ R_2 } , j } 
\eea 

The construction of gauge invariant operators is 
done as 
$ tr_{ m +  n } 
( Q^{ R }_{ R_1 , R_2 ; i , j }  ( \bX \otimes \bX^{ \dagger }) ) $.  
Restricted schur operators  were 
 investigated in \cite{Bhattacharyya:2008xy}. 
A similar formula to (\ref{inverseQ}) was given.

%%%%%%%%%%%%%%%%%%%%%%%%%%%%%%%%%%%%%%%%%%%%%%%%%%%%%%%%%%%%%%%%%%%%%%%%%

\section{Vershik-Okounkov approach}
\label{sec:vershikokounkov}
In this section, we review the Vershik-Okounkov 
approach \cite{VershikOkounkov}
to the representation theory of the symmetric group, 
which we use in section \ref{sec:examples}.

For the group algebra 
of the symmetric group $S_{n}$, 
define 
\begin{eqnarray}
X_{i}=(1,i)+(2,i)+\cdots +(i-1,i)
\end{eqnarray}
for $i=2,\cdots,n$ and $X_{1}=0$. 
This set gives us maximally commuting elements of the group algebra. 
They are called the Young-Jucys-Murphy elements (YJM-elements). 

In every irreducible representation, 
the eigenvalues of the $X_i$ uniquely choose a basis.  
Let $v$ be a basis vector in an irrep. We denote the
vector of  eigenvalues by  
\begin{eqnarray}
\alpha(v)=(a_{1},a_{2},\cdots,a_{n})
\end{eqnarray}
where $a_{i}$ is  the eigenvalue of $X_{i}$ on $v$ 
($a_{1}=0$ due to $X_{1}=0$).
It is a non-trivial result that the 
$a_{i}$ can be read off from standard Young tableaux. 
Indeed $a_i$ is equal to the content  of a box containing the  number $i$ 
in a standard Young tableau. 
The content of a box is 
its $x$-coordinate minus its $y$-coordinate, where 
$x$-axis is drawn from left to right and $y$-axis 
is from top to bottom. 
The content is shown in figure \ref{content}. 

\begin{figure}[htb]
\begin{center}
\setlength{\unitlength}{1mm}
\hspace{1cm}
\begin{picture}(25,18)(0,-18)
\put(0,0){\line(1,0){24}}
\put(0,-6){\line(1,0){24}}
\put(0,-12){\line(1,0){18}}
\put(0,-18){\line(1,0){6}}
\put(0,0){\line(0,-1){18}}
\put(6,0){\line(0,-1){18}}
\put(12,0){\line(0,-1){12}}
\put(18,0){\line(0,-1){12}}
\put(24,0){\line(0,-1){6}}

\put(0,-6){\makebox(6,6){$0$}}
\put(6,-6){\makebox(6,6){$1$}}
\put(0,-12){\makebox(6,6){$-1$}}
\put(0,-18){\makebox(6,6){$-2$}}
\put(6,-12){\makebox(6,6){$0$}}
\put(12,-12){\makebox(6,6){$1$}}
\put(12,-6){\makebox(6,6){$2$}}
\put(18,-6){\makebox(6,6){$3$}}
\put(26,-6){\makebox(6,6){$\cdots$}}
\put(0,-24.5){\makebox(6,6){$\vdots$}}
\end{picture}
\end{center}
\caption{Contents of boxes in Young tableau}
\label{content}
\end{figure}

Le us show some examples. 
The first example is the $[2,1]$ representation of $S_{3}$. 
As shown in figure \ref{S3standard}, 
there are two standard tableaux, which means the dimension 
of this representation is $2$. 
For a representation, 
$X_{i}$ are given by 
\begin{eqnarray}
&&
X_{2}=\mbox{diag}(-1,1) \cr
&&
X_{3}=\mbox{diag}(1,-1) 
\label{symmetricaodg21S3}
\end{eqnarray}
Basis vectors $e_{1}=(1,0)^{T}$ and $e_{2}=(0,1)^{T}$ correspond to 
the first and the second standard tableaux shown in 
figure \ref{S3standard} respectively. 

\begin{figure}[htb]
\begin{center}
\setlength{\unitlength}{1mm}
\hspace{1cm}
\begin{picture}(42,12)(0,-12)
\put(0,0){\line(1,0){12}}
\put(0,-6){\line(1,0){12}}
\put(0,-12){\line(1,0){6}}
\put(0,0){\line(0,-1){12}}
\put(6,0){\line(0,-1){12}}
\put(12,0){\line(0,-1){6}}

\put(0,-6){\makebox(6,6){$1$}}
\put(6,-6){\makebox(6,6){$3$}}
\put(0,-12){\makebox(6,6){$2$}}

\put(30,0){\line(1,0){12}}
\put(30,-6){\line(1,0){12}}
\put(30,-12){\line(1,0){6}}
\put(30,0){\line(0,-1){12}}
\put(36,0){\line(0,-1){12}}
\put(42,0){\line(0,-1){6}}

\put(30,-6){\makebox(6,6){$1$}}
\put(36,-6){\makebox(6,6){$2$}}
\put(30,-12){\makebox(6,6){$3$}}
\end{picture}
\end{center}
\caption{two standard tableaux for $\lambda=[2,1]$ of $S_{3}$}
\label{S3standard}
\end{figure}

The second example is 
the $[3,2]$ representation of $S_{5}$. 
$X_{i}$ are given by 
\begin{eqnarray}
&&
X_{2}=\mbox{diag}(-1,1,-1,1,1) \cr
&&
X_{3}=\mbox{diag}(1,-1,1,-1,2) \cr
&&
X_{4}=\mbox{diag}(0,0,2,2,-1) \cr
&&
X_{5}=\mbox{diag}(2,2,0,0,0) 
\end{eqnarray}
Basis vectors $e_{i}$ ($i=1,\cdots,5$) 
correspond to the $i$-th tableau given in 
figure \ref{32S5standard}. 

\begin{figure}[htb]
\begin{center}
\setlength{\unitlength}{1mm}
\hspace{1cm}
\begin{picture}(20,14)(0,-12)
\put(0,0){\line(1,0){18}}
\put(0,-6){\line(1,0){18}}
\put(0,-12){\line(1,0){12}}
\put(0,0){\line(0,-1){12}}
\put(6,0){\line(0,-1){12}}
\put(12,0){\line(0,-1){12}}
\put(18,0){\line(0,-1){6}}
\put(0,-6){\makebox(6,6){$1$}}
\put(6,-6){\makebox(6,6){$3$}}
\put(0,-12){\makebox(6,6){$2$}}
\put(6,-12){\makebox(6,6){$4$}}
\put(12,-6){\makebox(6,6){$5$}}
\end{picture}
%%%%%%%%
\begin{picture}(20,12)(0,-12)
\put(0,0){\line(1,0){18}}
\put(0,-6){\line(1,0){18}}
\put(0,-12){\line(1,0){12}}
\put(0,0){\line(0,-1){12}}
\put(6,0){\line(0,-1){12}}
\put(12,0){\line(0,-1){12}}
\put(18,0){\line(0,-1){6}}
\put(0,-6){\makebox(6,6){$1$}}
\put(6,-6){\makebox(6,6){$2$}}
\put(0,-12){\makebox(6,6){$3$}}
\put(6,-12){\makebox(6,6){$4$}}
\put(12,-6){\makebox(6,6){$5$}}
\end{picture}
%%%%%%%%%
\begin{picture}(20,12)(0,-12)
\put(0,0){\line(1,0){18}}
\put(0,-6){\line(1,0){18}}
\put(0,-12){\line(1,0){12}}
\put(0,0){\line(0,-1){12}}
\put(6,0){\line(0,-1){12}}
\put(12,0){\line(0,-1){12}}
\put(18,0){\line(0,-1){6}}
\put(0,-6){\makebox(6,6){$1$}}
\put(6,-6){\makebox(6,6){$3$}}
\put(0,-12){\makebox(6,6){$2$}}
\put(6,-12){\makebox(6,6){$5$}}
\put(12,-6){\makebox(6,6){$4$}}
\end{picture}
%%%%%%%%%
\begin{picture}(20,12)(0,-12)
\put(0,0){\line(1,0){18}}
\put(0,-6){\line(1,0){18}}
\put(0,-12){\line(1,0){12}}
\put(0,0){\line(0,-1){12}}
\put(6,0){\line(0,-1){12}}
\put(12,0){\line(0,-1){12}}
\put(18,0){\line(0,-1){6}}
\put(0,-6){\makebox(6,6){$1$}}
\put(6,-6){\makebox(6,6){$2$}}
\put(0,-12){\makebox(6,6){$3$}}
\put(6,-12){\makebox(6,6){$5$}}
\put(12,-6){\makebox(6,6){$4$}}
\end{picture}
%%%%%%%%%
\begin{picture}(20,12)(0,-12)
\put(0,0){\line(1,0){18}}
\put(0,-6){\line(1,0){18}}
\put(0,-12){\line(1,0){12}}
\put(0,0){\line(0,-1){12}}
\put(6,0){\line(0,-1){12}}
\put(12,0){\line(0,-1){12}}
\put(18,0){\line(0,-1){6}}
\put(0,-6){\makebox(6,6){$1$}}
\put(6,-6){\makebox(6,6){$2$}}
\put(0,-12){\makebox(6,6){$4$}}
\put(6,-12){\makebox(6,6){$5$}}
\put(12,-6){\makebox(6,6){$3$}}
\end{picture}
\end{center}
\caption{five standard tableaux for $\lambda=[3,2]$ of $S_{5}$}
\label{32S5standard}
\end{figure}

The last example is 
the $[2,2,1]$ representation of $S_{5}$. 
$X_{i}$ are given by 
\begin{eqnarray}
&&
X_{2}=\mbox{diag}(-1,-1,1,-1,1) \cr
&&
X_{3}=\mbox{diag}(-2,1,-1,1,-1) \cr
&&
X_{4}=\mbox{diag}(1,-2,-2,0,0) \cr
&&
X_{5}=\mbox{diag}(0,0,0,-2,-2) 
\end{eqnarray}
Basis vectors $e_{i}$ ($i=1,\cdots,5$) 
correspond to the $i$-th tableau given in 
figure \ref{221S5standard}.

\begin{figure}[htb]
\begin{center}
\setlength{\unitlength}{1mm}
\hspace{1cm}
\begin{picture}(20,20)(0,-18)
\put(0,0){\line(1,0){12}}
\put(0,-6){\line(1,0){12}}
\put(0,-12){\line(1,0){12}}
\put(0,-18){\line(1,0){6}}
\put(0,0){\line(0,-1){18}}
\put(6,0){\line(0,-1){18}}
\put(12,0){\line(0,-1){12}}
\put(0,-6){\makebox(6,6){$1$}}
\put(6,-6){\makebox(6,6){$4$}}
\put(0,-12){\makebox(6,6){$2$}}
\put(6,-12){\makebox(6,6){$5$}}
\put(0,-18){\makebox(6,6){$3$}}
\end{picture}
%%%%%%%%
\begin{picture}(20,14)(0,-18)
\put(0,0){\line(1,0){12}}
\put(0,-6){\line(1,0){12}}
\put(0,-12){\line(1,0){12}}
\put(0,-18){\line(1,0){6}}
\put(0,0){\line(0,-1){18}}
\put(6,0){\line(0,-1){18}}
\put(12,0){\line(0,-1){12}}
\put(0,-6){\makebox(6,6){$1$}}
\put(6,-6){\makebox(6,6){$3$}}
\put(0,-12){\makebox(6,6){$2$}}
\put(6,-12){\makebox(6,6){$5$}}
\put(0,-18){\makebox(6,6){$4$}}
\end{picture}
%%%%%%%%%
\begin{picture}(20,14)(0,-18)
\put(0,0){\line(1,0){12}}
\put(0,-6){\line(1,0){12}}
\put(0,-12){\line(1,0){12}}
\put(0,-18){\line(1,0){6}}
\put(0,0){\line(0,-1){18}}
\put(6,0){\line(0,-1){18}}
\put(12,0){\line(0,-1){12}}
\put(0,-6){\makebox(6,6){$1$}}
\put(6,-6){\makebox(6,6){$2$}}
\put(0,-12){\makebox(6,6){$3$}}
\put(6,-12){\makebox(6,6){$5$}}
\put(0,-18){\makebox(6,6){$4$}}
\end{picture}
%%%%%%%%%
\begin{picture}(20,14)(0,-18)
\put(0,0){\line(1,0){12}}
\put(0,-6){\line(1,0){12}}
\put(0,-12){\line(1,0){12}}
\put(0,-18){\line(1,0){6}}
\put(0,0){\line(0,-1){18}}
\put(6,0){\line(0,-1){18}}
\put(12,0){\line(0,-1){12}}
\put(0,-6){\makebox(6,6){$1$}}
\put(6,-6){\makebox(6,6){$3$}}
\put(0,-12){\makebox(6,6){$2$}}
\put(6,-12){\makebox(6,6){$4$}}
\put(0,-18){\makebox(6,6){$5$}}
\end{picture}
%%%%%%%%%
\begin{picture}(20,14)(0,-18)
\put(0,0){\line(1,0){12}}
\put(0,-6){\line(1,0){12}}
\put(0,-12){\line(1,0){12}}
\put(0,-18){\line(1,0){6}}
\put(0,0){\line(0,-1){18}}
\put(6,0){\line(0,-1){18}}
\put(12,0){\line(0,-1){12}}
\put(0,-6){\makebox(6,6){$1$}}
\put(6,-6){\makebox(6,6){$2$}}
\put(0,-12){\makebox(6,6){$3$}}
\put(6,-12){\makebox(6,6){$4$}}
\put(0,-18){\makebox(6,6){$5$}}
\end{picture}
\end{center}
\caption{five standard tableaux for $\lambda=[2,2,1]$ of $S_{5}$}
\label{221S5standard}
\end{figure}

%%%%%%%%%%%%%%%%%%%%%%%%%%%%%%%%%%%%%%%%%%%%%%%%%%%%%%%%%%%%%%%%%%%%%%%%%

\section{Representation of Brauer algebra}
\label{sec:Brauerrep}
In this section, we review the construction of 
representations of the Brauer algebra $B_{N}(m,n)$ based on 
\cite{BCHLLS}. 

Before going to the Brauer algebra, we review the case of 
the symmetric group $S_{n}$. 
The irreducible representations of the symmetric group $S_{n}$ 
are indexed by partitions $\lambda$ with $n$ boxes. 
The dimension of the irreducible representation is equal to 
the number of 
standard tableaux of partition $\lambda$. 
A standard tableau is defined by 
filling integers $1,\cdots,n$ 
into the boxes such that 
the numbers in the tableau increase from 
left to right in each row and 
increase top to bottom down in each column. 
Some examples of standard tableaux 
are presented in figure \ref{S3standard} - \ref{221S5standard}.

We now construct a representation of $S_{n}$. 
The construction starts with defining a tensor in $V^{\otimes n}$
\begin{equation}
\beta_{\tau}=u_{1}\otimes \cdots \otimes u_{n}
\end{equation}
where $u_{k}=v_{j}$ if $k$ is in the $j$-th row of 
a standard tableau $\tau$. 
We next define 
$t_{\tau}=y_{\tau}\beta_{\tau}$ where 
$y_{\tau}$ is the Young symmetriser for $\tau$. 
This $t_{\tau}$ gives a set of linearly independent bases 
in the irreducible $S_{n}$-module. 
Concretely, let us construct the $[2,1]$ representation. 
We denote the two standard tableaux 
in figure \ref{S3standard} 
as $\tau_{1}$ and $\tau_{2}$. 
For the first tableau $\tau_{1}$, 
we have $\beta_{\tau_{1}}=v_{1}\otimes v_{2}\otimes v_{1}$
and 
\begin{eqnarray}
t_{\tau_{1}}=\frac{1}{4}(1-(12))(1+(13))\beta_{\tau_{1}}
=\frac{1}{2}\left(
v_{1}\otimes v_{2}\otimes v_{1}
-v_{2}\otimes v_{1}\otimes v_{1}
\right)
\end{eqnarray}
and for $\tau_{2}$, we have 
$\beta_{\tau_{2}}=v_{1}\otimes v_{1}\otimes v_{2}$
and 
\begin{eqnarray}
t_{\tau_{2}}=\frac{1}{4}(1-(13))(1+(12))\beta_{\tau_{2}}
=\frac{1}{2}\left(
v_{1}\otimes v_{1}\otimes v_{2}
-v_{2}\otimes v_{1}\otimes v_{1}
\right)
\end{eqnarray}
Acting with $(12)$ and $(13)$ on these states, we have  
$(12)t_{\tau_{1}}=-t_{\tau_{1}}$, 
$(12)t_{\tau_{2}}=t_{\tau_{2}}-t_{\tau_{1}}$, and 
$(13)t_{\tau_{1}}=t_{\tau_{2}}$, $(13)t_{\tau_{2}}=t_{\tau_{1}}$.  
Then we obtain the following representation matrices 
\begin{eqnarray}
(12)=
\left(
\begin{array}{ccc}
1 & 0 \\
-1& -1 \\
\end{array}
\right)
\quad 
(23)=
\left(
\begin{array}{ccc}
0 & 1 \\
1& 0 \\
\end{array}
\right) 
\label{repof21S3}
\end{eqnarray}
%Note that this expression is different from 
%(\ref{symmetricaodg21S3}). 

We next consider the Brauer algebra $B_{N}(m,n)$. 
The irreducible representation of $B_{N}(m,n)$ 
is indexed by a sequence of integers 
$\gamma=(\gamma_{1},\cdots,\gamma_{N})$
obeying $\gamma_{1}\ge\cdots \ge \gamma_{N}$. 
This sequence of integers is called $N$-staircase. 
Because the postive integers of the sequence 
determine a partition, 
a staircase contains two partitions 
$\gamma^{+}$ 
and 
$\gamma^{-}$,   
which 
come from the positive and negative parts of $\gamma$ respectively. 
If $\gamma^{+}$ and $\gamma^{-}$ are given by 
a partition of $m-k$ and that of $n-k$ respectively, 
choosing a staircase $\gamma$ is equivalent to 
choosing a set of $(\gamma^{+},\gamma^{-},k)$, where 
$k$ is an integer with $0\le k\le min(m,n)$. 
It follows from these definitions that 
$c_{1}(\gamma^{+})+c_{1}(\gamma^{-})\le N$, where 
$c_{1}(\gamma^{+})$ is the length of the first column of $\gamma^{+}$. 
In this section, we assume this condition is always satisfied.

We introduce two set of integers 
${\cal P}=\{1,2,\cdots,m\}$ and ${\cal Q}=\{m+1,\cdots,m+n\}$. 
Let $\underline{p}=(p_{1},\cdots,p_{k})$ and 
$\underline{q}=(q_{1},\cdots,q_{k})$ be ordered subsets 
of ${\cal P}$ and ${\cal Q}$, and 
$\underline{p}^{c}$ and $\underline{q}^{c}$ 
be complements of 
$\underline{p}$ and $\underline{q}$ in  ${\cal P}$ and ${\cal Q}$. 
We define a set 
$(\tau,\underline{p},\underline{q})$ 
of standard tableau $\tau$ and ordered subsets of integers 
$\underline{p}$ and $\underline{q}$. 
$\tau=[\tau^{+},\tau^{-}]$ is called standard 
if both $\tau^{+}$ and $\tau^{-}$ are standard. 
$\tau^{+}$ ($\tau^{-}$) contains numbers in 
$\underline{p}^{c}$ ($\underline{q}^{c}$). 
The dimension of irreducible representations of Brauer algebra 
is equivalent to the number of the set $(\tau,\underline{p},\underline{q})$.  
For example, 
in the irreducible representation 
$\gamma=(k=2,[1],[1])$ of $B(3,3)$, 
there are $18$ sets, which are given by 

\begin{eqnarray*}
&&
\tau=[1,4] \quad 
\mbox{with 
$\underline{p}=(2,3)$ and $\underline{q}=(5,6)$}\cr
&&
\tau=[1,4] \quad 
\mbox{with 
$\underline{p}=(2,3)$ and $\underline{q}=(6,5)$}\cr
&&
\tau=[1,5] \quad 
\mbox{with 
$\underline{p}=(2,3)$ and $\underline{q}=(4,6)$}\cr
&&
\tau=[1,5] \quad 
\mbox{with 
$\underline{p}=(2,3)$ and $\underline{q}=(6,4)$}\cr
&&
\tau=[1,6] \quad 
\mbox{with 
$\underline{p}=(2,3)$ and $\underline{q}=(4,5)$}\cr
&&
\tau=[1,6] \quad 
\mbox{with 
$\underline{p}=(2,3)$ and $\underline{q}=(5,4)$}\cr
&&
\tau=[2,4] \quad 
\mbox{with 
$\underline{p}=(1,3)$ and $\underline{q}=(5,6)$}\cr
&&
\tau=[2,4] \quad 
\mbox{with 
$\underline{p}=(1,3)$ and $\underline{q}=(6,5)$}\cr
&&
\tau=[2,5] \quad 
\mbox{with 
$\underline{p}=(1,3)$ and $\underline{q}=(4,6)$}\cr
&&
\tau=[2,5] \quad 
\mbox{with 
$\underline{p}=(1,3)$ and $\underline{q}=(6,4)$}\cr
&&
\tau=[2,6] \quad 
\mbox{with 
$\underline{p}=(1,3)$ and $\underline{q}=(4,5)$}\cr
&&
\tau=[2,6] \quad 
\mbox{with 
$\underline{p}=(1,3)$ and $\underline{q}=(5,4)$}\cr
&&
\tau=[3,4] \quad 
\mbox{with 
$\underline{p}=(1,2)$ and $\underline{q}=(5,6)$}\cr
&&
\tau=[3,4] \quad 
\mbox{with 
$\underline{p}=(1,2)$ and $\underline{q}=(6,5)$}\cr
&&
\tau=[3,5] \quad 
\mbox{with 
$\underline{p}=(1,2)$ and $\underline{q}=(4,6)$}\cr
&&
\tau=[3,5] \quad 
\mbox{with 
$\underline{p}=(1,2)$ and $\underline{q}=(6,4)$}\cr
&&
\tau=[3,6] \quad 
\mbox{with 
$\underline{p}=(1,2)$ and $\underline{q}=(4,5)$}\cr
&&
\tau=[3,6] \quad 
\mbox{with 
$\underline{p}=(1,2)$ and $\underline{q}=(5,4)$}\cr
\end{eqnarray*}
where we have introduced an abbreviation $\tau=[1,4]$, 
which means the following Young tableau
\begin{center}
\setlength{\unitlength}{1mm}
$\tau=\bigl[ 
\mbox{
\begin{picture}(5,5)(0,1)
\put(0,0){\line(1,0){4}}
\put(0,0){\line(0,1){4}}
\put(4,0){\line(0,1){4}}
\put(0,4){\line(1,0){4}}
\put(0,0){\makebox(4,4){$1$}}
\end{picture},
\begin{picture}(5,5)(0,1)
\put(0,0){\line(1,0){4}}
\put(0,0){\line(0,1){4}}
\put(4,0){\line(0,1){4}}
\put(0,4){\line(1,0){4}}
\put(0,0){\makebox(4,4){$4$}}
\end{picture}
}
\bigr]$
\end{center}

For a given set of $(\tau,\underline{p},\underline{q})$, 
linearly independent basis 
$t_{\tau,\underline{p},\underline{q}}$ 
are given by the following.  
\begin{equation}
\beta_{\tau,\underline{p},\underline{q}}
=u_{1}\otimes \cdots \otimes u_{m}\otimes 
u_{m+1}^{\ast}\otimes \cdots \otimes u_{m+n}^{\ast}
\end{equation}
is a tensor whose factors are defined by 
\begin{eqnarray*}
&&u_{l}=\left\{ \begin{array}{ll}
v_{1} & \quad \mbox{if $l \in \underline{p}$} \\
v_{j} & \quad \mbox{if $l \in \underline{p}^{c}$ and 
$l$ in the $j$-th row of $\tau^{+}$} \\
\end{array} \right. \cr
&&u_{l}^{\ast}=\left\{ \begin{array}{ll}
v_{1}^{\ast} & \quad \mbox{if $l \in \underline{q}$} \\
v_{N-j+1}^{\ast} & \quad \mbox{if $l \in \underline{q}^{c}$ and 
$l$ in the $j$-th row of $\tau^{-}$} \\
\end{array} \right. 
\end{eqnarray*} 
From this, $t_{\tau,\underline{p},\underline{q}}$
is defined by 
\begin{eqnarray}
t_{\tau,\underline{p},\underline{q}}
=y_{\tau}C_{\underline{p},\underline{q}}
\beta_{\tau,\underline{p},\underline{q}}
\end{eqnarray}
where 
\begin{eqnarray}
C_{\underline{p},\underline{q}}
=C_{p_{1},q_{1}}\cdots C_{p_{k},q_{k}}
\end{eqnarray}
is the product of $k$ contractions. 
$C_{p,q}$ is the contraction acting on the $p$-th factor 
and the $q$-th factor as 
$Cv_{i}\otimes v_{j}=\delta_{ij}\sum_{l}v_{l}\otimes v_{l}$. 
$y_{\tau}=y_{\tau^{+}}y_{\tau^{-}}$ is the Young symmetriser. 
For the set of $(\tau,\underline{p},\underline{q})$ 
of $\gamma=(k=2,[1],[1])$, 
the bases are given by 
\begin{eqnarray}
&&
t^{(1)}=C_{25}C_{36}\beta_{\tau[1,4]}\quad
t^{(2)}=C_{35}C_{26}\beta_{\tau[1,4]}\cr
&&
t^{(3)}=C_{24}C_{36}\beta_{\tau[1,5]}\quad
t^{(4)}=C_{34}C_{26}\beta_{\tau[1,5]}\cr
&&
t^{(5)}=C_{24}C_{35}\beta_{\tau[1,6]}\quad
t^{(6)}=C_{34}C_{25}\beta_{\tau[1,6]}\cr
&&
t^{(7)}=C_{15}C_{36}\beta_{\tau[2,4]}\quad
t^{(8)}=C_{16}C_{35}\beta_{\tau[2,4]}\cr
&&
t^{(9)}=C_{14}C_{36}\beta_{\tau[2,5]}\quad
t^{(10)}=C_{16}C_{34}\beta_{\tau[2,5]}\cr
&&
t^{(11)}=C_{14}C_{35}\beta_{\tau[2,6]}\quad
t^{(12)}=C_{15}C_{34}\beta_{\tau[2,6]}\cr
&&
t^{(13)}=C_{15}C_{26}\beta_{\tau[3,4]}\quad
t^{(14)}=C_{16}C_{25}\beta_{\tau[3,4]}\cr
&&
t^{(15)}=C_{14}C_{26}\beta_{\tau[3,5]}\quad
t^{(16)}=C_{16}C_{24}\beta_{\tau[3,5]}\cr
&&
t^{(17)}=C_{14}C_{25}\beta_{\tau[3,6]}\quad
t^{(18)}=C_{15}C_{24}\beta_{\tau[3,6]}
\end{eqnarray}

Some examples of actions with the permutations and 
the contractions on these states 
are 
\begin{eqnarray}
&&(12)t^{(1)}=C_{15}C_{36}\beta_{\tau[2,4]}=t^{(7)}\cr
&&C_{14}t^{(1)}=C_{14}C_{25}C_{36}\beta_{\tau[1,4]}
=C_{25}C_{36}C_{14}\beta_{\tau[1,4]}=0 \cr
&&C_{14}t^{(3)}=C_{14}C_{24}C_{36}\beta_{\tau[1,5]}
=(12)C_{24}C_{36}\beta_{\tau[1,5]}
=C_{14}C_{36}\beta_{\tau[2,5]}=t^{(9)}\cr
&&C_{14}t^{(9)}=C_{14}C_{14}C_{36}\beta_{\tau[2,5]}
=NC_{14}C_{36}\beta_{\tau[2,5]}=Nt^{(9)}
\end{eqnarray}
These are calculated using relations such as  
\begin{eqnarray}
&& C_{i \bar j  }   C_{ i \bar k } = C_{i \bar j } ( \bar j \bar k  ) 
                     = ( \bar j \bar k ) C_{i \bar k } \cr 
&& C_{i \bar j }  C_{ k \bar j } = C_{i \bar j } ( ik ) 
= (ik)   C_{ k \bar j }
\cr
&&C_{i \bar j } C_{i \bar j } =NC_{i \bar j } 
\end{eqnarray}
Then representations for $(12)$, $(13)$, 
$(45)$ and $(46)$ 
are obtained as 
\begin{eqnarray}
&&(12)=
F_{1,7}+F_{2,8}+F_{3,9}
+F_{4,10}+F_{5,11}+F_{6,12}
+F_{13,14}+F_{15,16}+F_{17,18}\cr
&&
(13)=
F_{1,14}+F_{2,13}+F_{3,16}
+F_{4,15}+F_{5,18}+F_{6,17}
+F_{7,8}+F_{9,10}+F_{11,12} \cr
&&(45)=
F_{1,3}+F_{2,4}+F_{5,6}
+F_{7,9}+F_{8,10}+F_{11,12}
+F_{13,15}+F_{14,16}+F_{17,18}\cr
&&(46)=
F_{1,6}+F_{2,5}+F_{3,4}
+F_{7,12}+F_{8,11}+F_{9,10}
+F_{13,18}+F_{14,17}+F_{15,16}
\end{eqnarray}
Here we have introduced $F_{i,j}=E_{i,j}+E_{j,i}$, where 
$E_{i,j}$ is a matrix whose nonzero component is only $(i,j)$ 
component. 
$(23)$ and $(56)$ can be computed 
by $(23)=(12)(13)(12)$ and 
$(56)=(45)(46)(45)$. 

The contraction $C_{14}$ is calculated as 
\begin{eqnarray}
C_{14}=E_{9,3}
+E_{15,4}+E_{11,5}+E_{17,6}+E_{9,7}+E_{11,8}+NE_{9,9}+E_{9,10}
+NE_{11,11} \cr
+E_{11,12}
+E_{15,13}
+E_{17,14}
+NE_{15,15}
+E_{15,16}
+NE_{17,17}
+E_{17,18}
\end{eqnarray}
Other contractions such as $C_{15}$ or $C_{26}$ 
can be calculated by 
\begin{eqnarray}
C_{i j } = (i1) (4j ) C_{1 4 }  (i1) (4j ) 
\end{eqnarray}
where $i=1,2,3$ and $j=4,5,6$.

%%%%%%%%%%%%%%%%%%%%%%%%%%%%%%%%%


\begin{thebibliography}{99}


\bibitem{malda}
  J.~M.~Maldacena,
  ``The large N limit of superconformal field theories and supergravity,''
  Adv.\ Theor.\ Math.\ Phys.\  {\bf 2} (1998) 231
  [Int.\ J.\ Theor.\ Phys.\  {\bf 38} (1999) 1113]
  [arXiv:hep-th/9711200].
  %%CITATION = IJTPB,38,1113;%%

\bibitem{gkp}
  S.~S.~Gubser, I.~R.~Klebanov and A.~M.~Polyakov,
  ``Gauge theory correlators from non-critical string theory,''
  Phys.\ Lett.\  B {\bf 428} (1998) 105
  [arXiv:hep-th/9802109].
  %%CITATION = PHLTA,B428,105;%%

\bibitem{wit}
  E.~Witten,
  ``Anti-de Sitter space and holography,''
  Adv.\ Theor.\ Math.\ Phys.\  {\bf 2} (1998) 253
  [arXiv:hep-th/9802150].
  %%CITATION = 00203,2,253;%%


\bibitem{cjr}
  S.~Corley, A.~Jevicki and S.~Ramgoolam,
  ``Exact correlators of giant gravitons from dual N = 4 SYM theory,''
  Adv.\ Theor.\ Math.\ Phys.\  {\bf 5} (2002) 809
  [arXiv:hep-th/0111222].
  %%CITATION = 00203,5,809;%%

\bibitem{ber} 
D. Berenstein, 
``A toy model for the AdS/CFT correspondence,''
JHEP 0407 (2004) 018 
[arXiv:hep-th/0403110].

\bibitem{mst}
J. McGreevy, L. Susskind and  N. Toumbas, 
``Invasion of the Giant Gravitons from Anti-de Sitter Space,'' 
JHEP 0006 (2000) 008 [arXiv:hep-th/0003075].

\bibitem{llm} 
H. Lin, O. Lunin and J. Maldacena, 
``Bubbling AdS space and 1/2 BPS geometries,''
JHEP 0410 (2004) 025 
[arXiv:hep-th/0409174].

\bibitem{bbjs}
  V.~Balasubramanian, J.~de Boer, V.~Jejjala and J.~Simon,
  ``The library of Babel: On the origin of gravitational thermodynamics,''
  JHEP {\bf 0512} (2005) 006
  [arXiv:hep-th/0508023].
  %%CITATION = JHEPA,0512,006;%%

\bibitem{kr}
  Y.~Kimura and S.~Ramgoolam,
  ``Branes, Anti-Branes and Brauer Algebras in Gauge-Gravity duality,''
  JHEP {\bf 0711} (2007) 078
  [arXiv:0709.2158 [hep-th]].
  %%CITATION = JHEPA,0711,078;%%
 

\bibitem{bhr}
  T.~W.~Brown, P.~J.~Heslop and S.~Ramgoolam,
  ``Diagonal multi-matrix correlators and BPS operators in N=4 SYM,''
  JHEP {\bf 0802} (2008) 030
  [arXiv:0711.0176 [hep-th]].
  %%CITATION = JHEPA,0802,030;%%

\bibitem{bcd}
  R.~Bhattacharyya, S.~Collins and R.~d.~M.~Koch,
  ``Exact Multi-Matrix Correlators,''
  JHEP {\bf 0803} (2008) 044
  [arXiv:0801.2061 [hep-th]].
  %%CITATION = JHEPA,0803,044;%%


\bibitem{bhr2}
  T.~W.~Brown, P.~J.~Heslop and S.~Ramgoolam,
  ``Diagonal free field matrix correlators, global symmetries and giant
  gravitons,''
  arXiv:0806.1911 [hep-th].
  %%CITATION = ARXIV:0806.1911;%%

\bibitem{cr} 
  S.~Corley and S.~Ramgoolam,
  ``Finite factorization equations and sum rules for BPS correlators in  N = 4
  SYM theory,''
  Nucl.\ Phys.\  B {\bf 641} (2002) 131
  [arXiv:hep-th/0205221].
  %%CITATION = NUPHA,B641,131;%%

\bibitem{swreview} 
  S.~Ramgoolam,
  ``Schur-Weyl duality as an instrument of Gauge-String duality,''
  arXiv:0804.2764 [hep-th].
  %%CITATION = ARXIV:0804.2764;%%



\bibitem{integinfo}
  V.~Balasubramanian, B.~Czech, K.~Larjo and J.~Simon,
  ``Integrability vs. information loss: A simple example,''
  JHEP {\bf 0611} (2006) 001
  [arXiv:hep-th/0602263].
  %%CITATION = JHEPA,0611,001;%%


\bibitem{halverson} 
T. Halverson, ``Characters of the centralizer algebras of mixed tensor
representations of $GL(r,\C ) $ and the quantum group $U_q ( GL(r , \C ) $, ''
Pacific Journal of Mathematics, Vol. 174, No. 2 , 1996. 


\bibitem{goodwall} 
  R. Goodman and N. Wallach, ``Representations and Invariants of 
  classical groups,'' CUP 1998.

\bibitem{hassbut} 
R W Haas and P H Butler, ``Symmetric and Unitary group representations : 
I Duality theory, '' J. Phys. A , Math. Gen. {\bf 17} (1984) 61-74. 

\bibitem{itzub} 
C.~Itzykson and B.~Zuber,
``Quantum Field Theory,''  McGraw-Hill Inc. 1980.


\bibitem{GT} 
  D.~J.~Gross and W.~Taylor,
  ``Two-dimensional QCD is a string theory,''
  Nucl.\ Phys.\  B {\bf 400} (1993) 181
  [arXiv:hep-th/9301068].
  %%CITATION = NUPHA,B400,181;%%

%\cite{Kimura:2008gs}
\bibitem{Kimura:2008gs}
  Y.~Kimura and S.~Ramgoolam,
  ``Holomorphic maps and the complete 1/N expansion of 2D SU(N) Yang-Mills,''
  JHEP {\bf 0806} (2008) 015
  [arXiv:0802.3662 [hep-th]].
  %%CITATION = JHEPA,0806,015;%%

\bibitem{dolnapwit} 
  L.~Dolan, C.~R.~Nappi and E.~Witten,
  ``A relation between approaches to integrability in superconformal
  Yang-Mills theory,''
  JHEP {\bf 0310} (2003) 017
  [arXiv:hep-th/0308089].
  %%CITATION = JHEPA,0310,017;%%


\bibitem{cmr}
  S.~Cordes, G.~W.~Moore and S.~Ramgoolam,
  ``Lectures On 2-D Yang-Mills Theory, Equivariant Cohomology And Topological
  Field Theories,''
  Nucl.\ Phys.\ Proc.\ Suppl.\  {\bf 41} (1995) 184
  [arXiv:hep-th/9411210].
  %%CITATION = NUPHZ,41,184;%%

\bibitem{tom1loop}
  T.~W.~Brown,
  ``Permutations and the Loop,''
  JHEP {\bf 0806} (2008) 008
  [arXiv:0801.2094 [hep-th]].
  %%CITATION = JHEPA,0806,008;%%



\bibitem{martin} 
A.~Cox, M.~d.~Visscher, S.~Doty, P.~Martin, 
``On the Blocks of the walled Brauer algebra,'' 
 [arXiv:0709.0851].

\bibitem{malstrom}
  J.~M.~Maldacena and A.~Strominger,
  ``AdS(3) black holes and a stringy exclusion principle,''
  JHEP {\bf 9812} (1998) 005
  [arXiv:hep-th/9804085].
  %%CITATION = JHEPA,9812,005;%%

\bibitem{jevram}
  A.~Jevicki and S.~Ramgoolam,
  ``Non-commutative gravity from the AdS/CFT correspondence,''
  JHEP {\bf 9904} (1999) 032
  [arXiv:hep-th/9902059].
  %%CITATION = JHEPA,9904,032;%%

\bibitem{horam}
  P.~M.~Ho, S.~Ramgoolam and R.~Tatar,
  ``Quantum spacetimes and finite N effects
 in 4D super Yang-Mills  theories,''
  Nucl.\ Phys.\  B {\bf 573} (2000) 364
  [arXiv:hep-th/9907145].
  %%CITATION = NUPHA,B573,364;%%

\bibitem{DVV}
  R.~Dijkgraaf, E.~P.~Verlinde and H.~L.~Verlinde,
  ``Matrix string theory,''
  Nucl.\ Phys.\  B {\bf 500} (1997) 43
  [arXiv:hep-th/9703030].
  %%CITATION = NUPHA,B500,43;%%

\bibitem{finiteal}
  G.~W.~Moore,
  ``Finite In All Directions,''
  arXiv:hep-th/9305139.
  %%CITATION = HEP-TH/9305139;%%

\bibitem{gopak}
  R.~Gopakumar,
  ``From free fields to AdS,''
  Phys.\ Rev.\  D {\bf 70} (2004) 025009
  [arXiv:hep-th/0308184].
  %%CITATION = PHRVA,D70,025009;%%

\bibitem{hagsun}
  P.~Haggi-Mani and B.~Sundborg,
  ``Free large N supersymmetric Yang-Mills theory as a string theory,''
  JHEP {\bf 0004} (2000) 031
  [arXiv:hep-th/0002189].
  %%CITATION = JHEPA,0004,031;%%

\bibitem{ramthesis}
A.~Ram, 
Thesis, Univ. Cal. San Diego 1991, Chapter 1. 

\bibitem{bbfh}
  V.~Balasubramanian, D.~Berenstein, B.~Feng and M.~x.~Huang,
  ``D-branes in Yang-Mills theory and emergent gauge symmetry,''
  JHEP {\bf 0503} (2005) 006
  [arXiv:hep-th/0411205].
  %%CITATION = JHEPA,0503,006;%%

\bibitem{robgtsti}
  R.~de Mello Koch, J.~Smolic and M.~Smolic,
  ``Giant Gravitons - with Strings Attached (I),''
  JHEP {\bf 0706} (2007) 074
  [arXiv:hep-th/0701066].
  %%CITATION = JHEPA,0706,074;%%

\bibitem{robgtstii}
  R.~de Mello Koch, J.~Smolic and M.~Smolic,
  ``Giant Gravitons - with Strings Attached (II),''
  JHEP {\bf 0709} (2007) 049
  [arXiv:hep-th/0701067].
  %%CITATION = JHEPA,0709,049;%%

\bibitem{robgtstiii}
  D.~Bekker, R.~de Mello Koch and M.~Stephanou,
  ``Giant Gravitons - with Strings Attached (III),''
  JHEP {\bf 0802} (2008) 029
  [arXiv:0710.5372 [hep-th]].
  %%CITATION = JHEPA,0802,029;%%

\bibitem{Bhattacharyya:2008xy}
  R.~Bhattacharyya, R.~de Mello Koch and M.~Stephanou,
  ``Exact Multi-Restricted Schur Polynomial Correlators,''
  JHEP {\bf 0806} (2008) 101
  [arXiv:0805.3025 [hep-th]].
  %%CITATION = JHEPA,0806,101;%%


\bibitem{VershikOkounkov}
A.~M~.Vershik and A.~Yu.~Okounkov, 
``A New Approach to the Representation Thoery 
of the Symmetric Groups. II,'' 
Zapiski Seminarod POMI (In Russian) v.307, 2004, 
math/0503040.

\bibitem{BCHLLS} 
M.~Benkart, M.~Chakrabarti, T.~Halverson, C.~Lee, R.~Leduc and J.~Stroomer, 
``Tensor product representations of general linear groups 
and their connections with Brauer algebras,'' 
J. Algebra, 166 (1994), 529.


\end{thebibliography}
\end{document}